\documentclass[apj]{emulateapj}

\def\ltsima{$\; \buildrel < \over \sim \;$}
\def\simlt{\lower.5ex\hbox{\ltsima}}
\def\gtsima{$\; \buildrel > \over \sim \;$}
\def\simgt{\lower.5ex\hbox{\gtsima}}
\def\kpc{{\rm\,kpc}}

\def\msun{{\rm\,M_\odot}}

\def \magnitude {$^{\rm m}$}

\def\deg{^\circ}

\def\s{\ifmmode \widetilde \else \~\fi}
\def\={\overline}

\def\spose#1{\hbox to 0pt{#1\hss}}

\def\eg{{ e.g.,\ }}
\def\ie{{ i.e.,\ }}
\def\lta{\mathrel{\spose{\lower 3pt\hbox{$\mathchar"218$}}
     \raise 2.0pt\hbox{$\mathchar"13C$}}}
\def\gta{\mathrel{\spose{\lower 3pt\hbox{$\mathchar"218$}}
     \raise 2.0pt\hbox{$\mathchar"13E$}}}
\def\Dt{\spose{\raise 1.5ex\hbox{\hskip3pt$\mathchar"201$}}}    
\def\dt{\spose{\raise 1.0ex\hbox{\hskip2pt$\mathchar"201$}}}    

\def\dotsfill{\leaders\hbox to 1em{\hss.\hss}\hfill}
\def\sun{\odot}

\def\FeH{{\rm[Fe/H]}}

\shorttitle{New dwarf galaxies in the surroundings of M31}
\shortauthors{J. C. Richardson et al.}

\begin{document}


\title{PAndAS' progeny: extending the M31 dwarf galaxy cabal.$^{9}$}


\author{Jenny C. Richardson$^1$, Mike Irwin$^1$, Alan W. McConnachie$^2$, Nicolas F. Martin$^3$, Aaron Dotter$^2$, Annette M. N. Ferguson$^4$, Rodrigo A. Ibata$^5$, Scott Chapman $^1$, Geraint F. Lewis$^{6}$, Nial R. Tanvir$^{7}$, R. Michael Rich$^{8}$}
\email{jcr@ast.cam.ac.uk}

\altaffiltext{1}{Institute of Astronomy, Madingley Road, Cambridge,
CB3 0HA, U.K.}  
\altaffiltext{2}{NRC Herzberg Institute for Astrophysics, 5071 West
Saanich Road, Victoria, British Columbia, Canada, V9E 2E7}
\altaffiltext{3}{Max-Planck-Institut f\"ur Astronomie, K\"onigstuhl
17, D-69117 Heidelberg, Germany}
\altaffiltext{4}{Institute for Astronomy, University of Edinburgh,
Royal Observatory, Blackford Hill, Edinburgh, EH9 3HJ, U.K.}
\altaffiltext{5}{Observatoire de Strasbourg, 11, rue de
l'Universit\'e, F-67000, Strasbourg, France}
\altaffiltext{6}{Institute of Astronomy, School of Physics, University
of Sydney, NSW 2006, Australia.}
\altaffiltext{7}{Department of Physics and Astronomy, University of
Leicester, University Road, Leicester LE1 7RH, U.K.}
\altaffiltext{8}{Physics and Astronomy Building, 430 Portola Plaza,
Box 951547, Department of Physics and Astronomy,University of
California, Los Angeles, CA 90095-1547, USA.}

\begin{abstract}

We present the discovery of five new dwarf galaxies,
Andromeda~XXIII--XXVII, located in the outer halo of M31. These
galaxies were discovered during the second year of data from the
Pan-Andromeda Archaeological Survey (PAndAS), a photometric survey of
the M31/M33 subgroup conducted with the MegaPrime/MegaCam wide-field
camera on the Canada-France-Hawaii Telescope. The current PAndAS
survey now provides an almost complete panoramic view of the M31 halo
out to an average projected radius of $\sim150\kpc$. Here we present
for the first time the metal-poor stellar density map for this whole
region, not only as an illustration of the discovery space for satellite 
galaxies, but also as a birds-eye view of the ongoing assembly process of 
an L$_*$ disk galaxy.

Four of the newly discovered satellites appear as well-defined spatial
over-densities of stars lying on the expected locus of metal-poor
($-2.5 < \FeH < -1.3$) red giant branch stars at the distance of M31.
The fifth over-density, And~XXVII, is embedded in an extensive stream
of such stars and is possibly the remnant of a strong tidal disruption
event.  Based on distance estimates from horizontal branch magnitudes,
all five have metallicities typical of dwarf spheroidal galaxies ranging
from [Fe/H] $= -1.7 \pm 0.2$ to [Fe/H] $= -1.9 \pm 0.2$ and absolute 
magnitudes ranging from $\rm{M}_V = -7.1 \pm 0.5$ to $\rm{M}_V = -10.2 \pm 0.5$.
These five additional satellites bring the number of dwarf spheroidal
galaxies in this region to twenty five and continue the trend whereby
the brighter dwarf spheroidal satellites of M31 generally have much
larger half-light radii than their Milky Way counterparts.

With an extended sample of M31 satellite galaxies we also revisit the spatial
distribution of this population and in particular we find that, within
the current projected limits of the PAndAS survey the surface density
of satellites is essentially constant out to 150 kpc.  This corresponds to a
radial density distribution of satellites varying as $r^{-1}$ a result 
seemingly in conflict with the predictions of cosmological simulations.

\end{abstract}


\section{Introduction}
\footnotetext[9]{Based on observations obtained with
MegaPrime/MegaCam, a joint project of CFHT and CEA/DAPNIA, at the
Canada-France-Hawaii Telescope (CFHT) which is operated by the
National Research Council (NRC) of Canada, the Institut National des
Science de l'Univers of the Centre National de la Recherche
Scientifique (CNRS) of France, and the University of Hawaii.}

The last decade has seen tremendous advances in our understanding of
our nearest giant neighbour, M31. A particular highlight has been the
first detailed exploration of the outer halo of the galaxy. The Isaac
Newton Telescope wide-field imaging survey of M31 mapped a $\sim
100\times 100$~kpc$^2$ region down to $\approx3$ magnitudes below the
red giant branch (RGB) tip and revealed a strikingly complex picture
of an inner halo replete with stellar substructure \citep{ibata01a,
ferguson02, irwin05}, and led to the discovery of Andromeda~XVII
\citep{irwin08}. With no sign of the substructure abating at the
survey edge, attention naturally turned to the more remote regions of
the M31 halo.  Using MegaCam on the Canada-France-Hawaii Telescope
(CFHT), \citealt{ibata07} mapped the stellar distribution in the
entire south-eastern quadrant in the radial range $\sim 30-150$~kpc,
with an extension towards M33. A rich array of tidal streams and
overdensities, globular clusters and satellite galaxies were revealed.
Encouraged by the success of these surveys we embarked on the
Pan-Andromeda Archaeological Survey (PAndAS, \citealt{mcconnachie09}),
a Large Program using the CFHT MegaCam imager to map the entire
stellar halos of both M31 and M33 out to distances of $\sim$150 and
$\sim$50 kpc respectively.

As the data came in, quadrant by quadrant, increasing numbers of new
dwarf spheroidal (dSph) satellite galaxies were revealed in the
outskirts of M31: And~XI, XII, XIII, XV, XVI, XVIII, XIX, XX, XXI
and~XXII \citep{martin06,ibata07,mcconnachie08,martin09}. In parallel
efforts, the existence of And~IX, X and XIV was discovered by other
groups \citep{zucker04, zucker07, majewski07}.  The recently acquired
second year of data for the PAndAS survey now yields an almost
complete perspective of the outer halo of M31 out to $\sim$150 kpc,
showing for the first time the full structural complexity of the
ongoing assembly process of an L$_*$ disk galaxy
(Fig.~\ref{fig_map}). Also clearly visible within this panoramic view
are multiple dwarf galactic satellites in the M31 system including
five new dwarf galaxies, And~XXIII, And~XXIV, And~XXV, And~XXVI and
And~XXVII, revealed here for the first time (circled in red in
Fig.~\ref{fig_map}).


However, our knowledge of the satellite system of M31 is still
limited.  Current wide-area ground-based photometric observations of
satellites at the distance of M31 cannot easily reach much deeper than
the horizontal branch, limiting detection to objects at the bright end
of the satellite luminosity function (M$_V \lesssim -6.5$). If M31 had
a population of ultra-faint dwarfs like the MW, they would not be
picked up in present surveys.  Spectroscopic observations are
currently even more constrained and are limited in practice to the
handful of brighter red giant branch RGB stars close to the tip of the
RGB (TRGB) or, if present, any bright asymptotic giant branch (AGB)
stars.  Despite these limitations much can already be said of the
generic properties of these new objects from the current photometric
survey data alone.

The absolute magnitudes of the recently discovered M31 satellites
range from $M_V=-6.4$ for And~XII and And~XX
\citep{martin06,mcconnachie08} to a surprisingly bright $M_V\lta-9.7$
for And~XVIII \citep{mcconnachie08}, patently showing the
incompleteness of the M31 satellite luminosity function in regions
that have so far only been surveyed with photographic plates.  We note
here that, even with hindsight, only one of these new discoveries
(AndXVIII) is visible on the earlier photographic sky surveys and none
of the five new satellites presented here are visible on earlier sky
surveys either. With their relative faintness and significant sizes
(half-light radii $>$~100pc), these new systems are usually assumed to
be (dSph) galaxies, \ie devoid of any significant amount of gas and
without significant recent star formation.  The absence of gas is
currently consistent with the results of \textsc{Hi} surveys but the
upper limits on their \textsc{Hi} content generally remains relatively
high, $2-3\times10^5\msun$ \citep{grcevich09}, though in some cases
the \textsc{Hi} limit is much tighter.  For example, \cite{chapman07}
constrained the \textsc{Hi} mass of And~XII to be $< 3
\times10^3\msun$.  In general, however, it is currently not possible
to rule out entirely that some of these galaxies may still contain
non-negligible amounts of gas.

The properties of the surviving satellite populations of L$_*$
galaxies, such as M31 and the MW, are presumably directly linked to
the overall evolution and assembly of the host systems.  Since M31 and
the MW evolved in a common environment, have similar morphological
types (SbI-II and SbcI-II), masses ($1-2 \times 10^{12}~M_{\sun}$,
\citealt{evans00}) and luminosities, one would expect their satellite
systems also to be similar. Intriguingly, a growing body of evidence
suggests that this is not the case. \cite{mcconnachie06a} compared the
structural properties of the then known MW and M31 dSph populations
and found systematic differences whereby the dSphs of M31 had
half-light radii typically twice as large as the MW dSph. As more
dSphs have been discovered around both galaxies, \cite{kalirai10} and
\cite{collins10} have used the kinematics of several M31 dSphs to show
that the brighter dSphs are systematically dynamically colder and have
lower central densities than their MW counterpart; while at the
fainter end of the dSph luminosity functions the structural properties
appear to overlap (\citealt{kalirai10}).  This potential
luminosity-dependent discrepancy between the observed properties of
the MW and M31 satellite systems is puzzling and as more satellites
are discovered is worth reassessing.

In this paper we analyze the properties of these new dwarfs and place
them in the context of the rest of the Andromedan satellite galaxy
population. Section~2 summarizes the PAndAS data and the processing
steps involved. Sections~3 and 4 present the new systems and their
structural properties and in section~5 we discuss them in the context
of the ensemble of Andromeda satellites and compare this to the MW
satellite system.

Throughout this paper, the distance moduli of M31 is assumed to be
$(m-M)_0=24.47\pm0.07$, or $785\pm25$ \citep{mcconnachie05}.

\section{The PAndAS survey}
PAndAS builds upon the previous CFHT/MegaCam surveys of M31 presented
in \citet{martin06}, \citet{ibata07} and \citet{mcconnachie08} and we
refer the reader to these papers for a full description of the
observing strategy, data reduction and data quality. The imaging was
carried out with the MegaPrime/MegaCam camera mounted on the
Canada-France-Hawaii Telescope which has 36 $2048\times4612$ CCDs with
a pixel size of 0.187\,arcsec and $1\times1$ deg$^2$ field of
view. Although the survey is comprised of contiguous imaging, small
gaps in exposures lead to a scientifically usable field-of-view of
$0.96\times0.94\,\textrm{deg}^2$ for each of the (currently 357)
pointings of the survey. Each field has been observed in the MegaCam
$g$ and $i$ filters for at least 1350\,s, split into $3\times450$\,s
dithered sub-exposures. Good seeing ($<0.8''$) and rigorous CFHT
on-site quality control ensures that the photometry reaches
$g\sim25.5$ and $i\sim24.5$ with a signal-to-noise ratio of 10. With
data of this quality the star/galaxy separation only begins to degrade
significantly at magnitudes fainter than this.

Data were preprocessed by the Elixir system at CFHT, this includes
de-biasing, flat-fielding and fringe-correcting the data as well as
determining the photometric zero-points. The data were then
transferred to Cambridge where they were further processed using a
version of the Cambridge Astronomical Survey Unit (CASU) photometry
pipeline \citep{irwin01} which was specially tailored to CFHT/MegaCam
observations.  Here the astrometry of individual frames are refined
and this information was used to register and then stack component
images to create the final image products from which the survey
catalogues are generated. Finally, the astrometry is further refined,
and objects from the catalogues are morphologically classified
(stellar, non-stellar, noise-like) before creating the final
band-merged $g$ and $i$ products.  The catalogues provide additional
quality control information and the classification step also computes
the aperture corrections required to place the photometry on an
absolute scale.  The band-merged catalogues for each field are then
combined to form an overall single entry $g$, $i$ catalogue for each
detected object. In this process objects lying within 1 arcsec of each
other are taken to be the same and the entry with the higher
signal-to-noise measure is retained.  Objects present only on $g$ or
$i$ are retained throughout this process.

In the following, unless otherwise stated, the magnitudes are
presented in their natural instrumental (AB) system without any
reddening correction (g, i).  When required for analysis, de-reddened
magnitudes ($g_0$ and $i_0$) or reddened model isochrones have been
determined from the \citet{schlegel98} $E(B-V)$ extinction maps, using
the following correction coefficients: $g_0=g-3.793E(B-V)$ and
$i_0=i-2.086E(B-V)$ listed in their Table~6.

\section{The PAndAS Metal-Poor Stellar Density Map}


\begin{figure*}
\begin{center}
\includegraphics[width=0.875\hsize,angle=270]{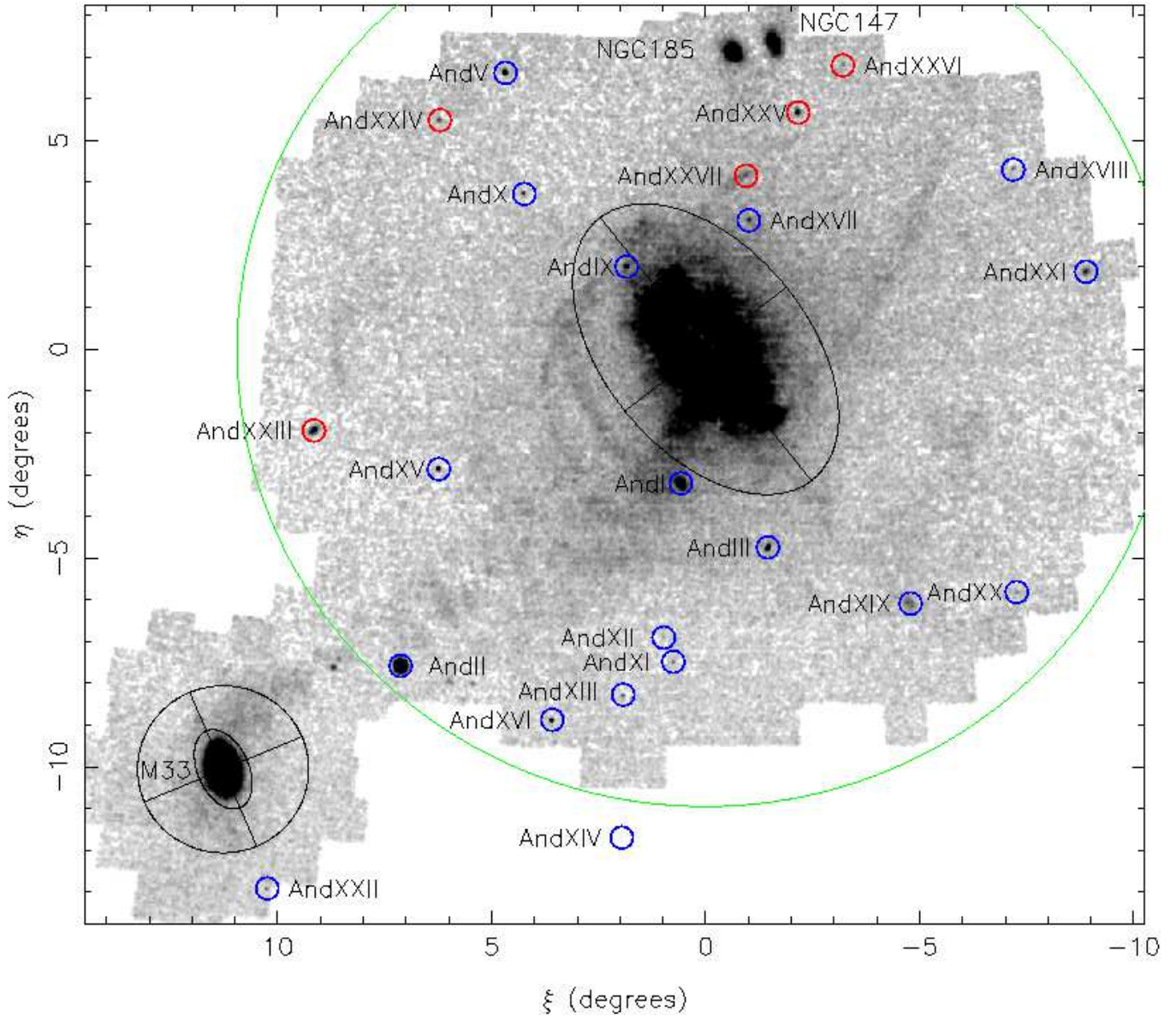}
\caption{\label{fig_map} Surface density map of stars with colors and
magnitudes consistent with belonging to metal-poor red giant branch
populations at the distance of M31.  The almost uniform underlying
background is mainly contributed by foreground stars in the Milky Way
together with a small residual contamination from unresolved compact
background galaxies.  All of the previously known M31 dwarf
spheroidals in this region covered by the survey are readily visible
as well-defined over-densities and are marked with blue circles. The
five new dwarf spheroidals are highlighted in red.  (And~XIV is the
dwarf spheroidal just south of the present survey area, while AndVI
and AndVII lie respectively well to the West and North of the region
shown.)  NGC147 and NGC185 appear at the top of the map and M33 at the
bottom left.  The green circle lies at a projected radius of 150 kpc
from the center of M31 within which most of the survey lies.  In
addition to the satellite galaxies numerous stellar streams and
substructures are visible.  Although the majority of small
over-densities are satellite galaxies of M31, a few to the southern
end of the map (not circled) are resolved globular cluster systems
picked out surrounding nearby low redshift background elliptical
galaxies.}
\end{center}
\end{figure*}

The spectacular results of the PAndAS survey are shown in
Fig.~\ref{fig_map}.  The map details the density distribution of all
candidate metal-poor stars in the stellar halos of M31 (M33) out to
150~kpc (50~kpc), covering more than 300~$\rm{deg}^2$, or around
$55,000~\rm{kpc}^2$ at the distance of M31.  Here stars are considered
as metal-poor candidates if they lie in the locus of the
Color-Magnitude Diagram (CMD) where RGB stars with metallicities of
$\approx -2.5 < \FeH < -1.3$ fall if at the systemic distance of M31
$\pm100$~kpc, and have $i$-band and $g$-band magnitudes satisfying
$21.0 < i < 24.0$ and $g < 25.0$.  In this we are implicitly assuming
that such stars belong to predominantly old ($\approx$10~Gyr) stellar
populations.

This is the first time that a deep ($\Sigma_V \simeq 33~\rm{mag
/arcsec}^2$) high resolution contiguous map of the majority of the
extended stellar halo of any L$_*$ galaxy has been observed. Since the
signatures of past accretions and mergers are preserved for longer
times in the outer regions of massive galaxies
(e.g. \citealt{johnston96}), the PAndAS survey has effectively
provided a unique insight into the accretion history of M31.  Although
the map presented in Fig.~\ref{fig_map} was optimized to search for
dSph-like structures at the distance of M31, it also clearly reveals
that M31 is looped by a series of giant stellar streams which have
been torn off from accreted satellite galaxies, as well as other more
diffuse stellar substructures. In particular, we note the great
stellar arc to the North-West seen here for the first time, the two
well-defined streams of stars crossing the South-East Minor axis and
the diffuse debris apparently emanating from M33 that is visible over
a large region indicative of a strong tidal interaction with M31 (see
\citealt{mcconnachie09}).  In addition numerous outer halo globular
clusters are also being found, many coincident with these stream-like
features \citep{mackey10}.

Interestingly, although part of the North-West arc was also seen by
\cite{tanaka10}, their stream ``F'', there is no evidence in our
metal-poor map for any over-density at the location of their stream
``E'', nor in any of our other maps made with different metallicity
cuts.

Our panoramic view also offers the most complete picture of the dwarf
satellite galaxy population of M31 to date. Previously known dwarf
spheroidals are circled in blue. The five new dwarf spheroidals, which
are the main subject of this paper, are circled in red.  Of the 25
currently known Andromedan dSphs, And~VI and VII lie well outside the
field-of-view shown here, while And~XIV lies just to the South of the
surveyed region and is shown with a blue circle.  Larger satellites of
M31 such as M32 and NGC205 are hidden within the dense inner halo, but
NGC147 and NGC185 make an appearance at the Northern edge of the
image. Including M31 and M33, but not counting the numerous streams
and other extended substructure, the number of galaxies in the
Andromedan system comfortably visible in the surveyed area in this
image totals 28.  Since these objects are distributed all over the
area surveyed with no obvious curtailment as a function of radius, it
is clear that even this large-scale panoramic view is still not large
enough to capture all the detail of the M31 outer halo systems.

\section{Andromedas XXIII - XXVII}
\label{desc}

\begin{figure}
\begin{center}
\includegraphics[width=0.44\hsize,angle=270]{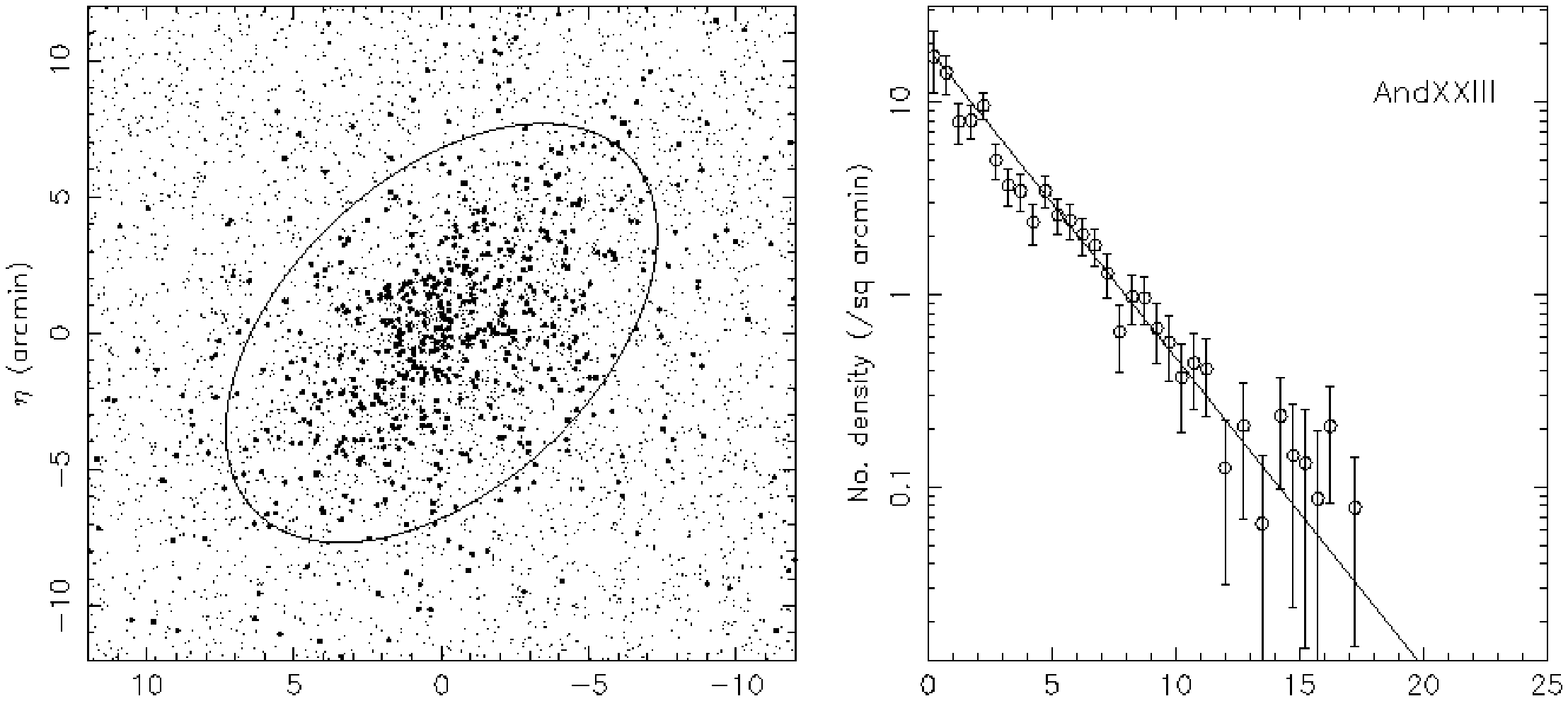}
\vskip0.1cm
\includegraphics[width=0.44\hsize,angle=270]{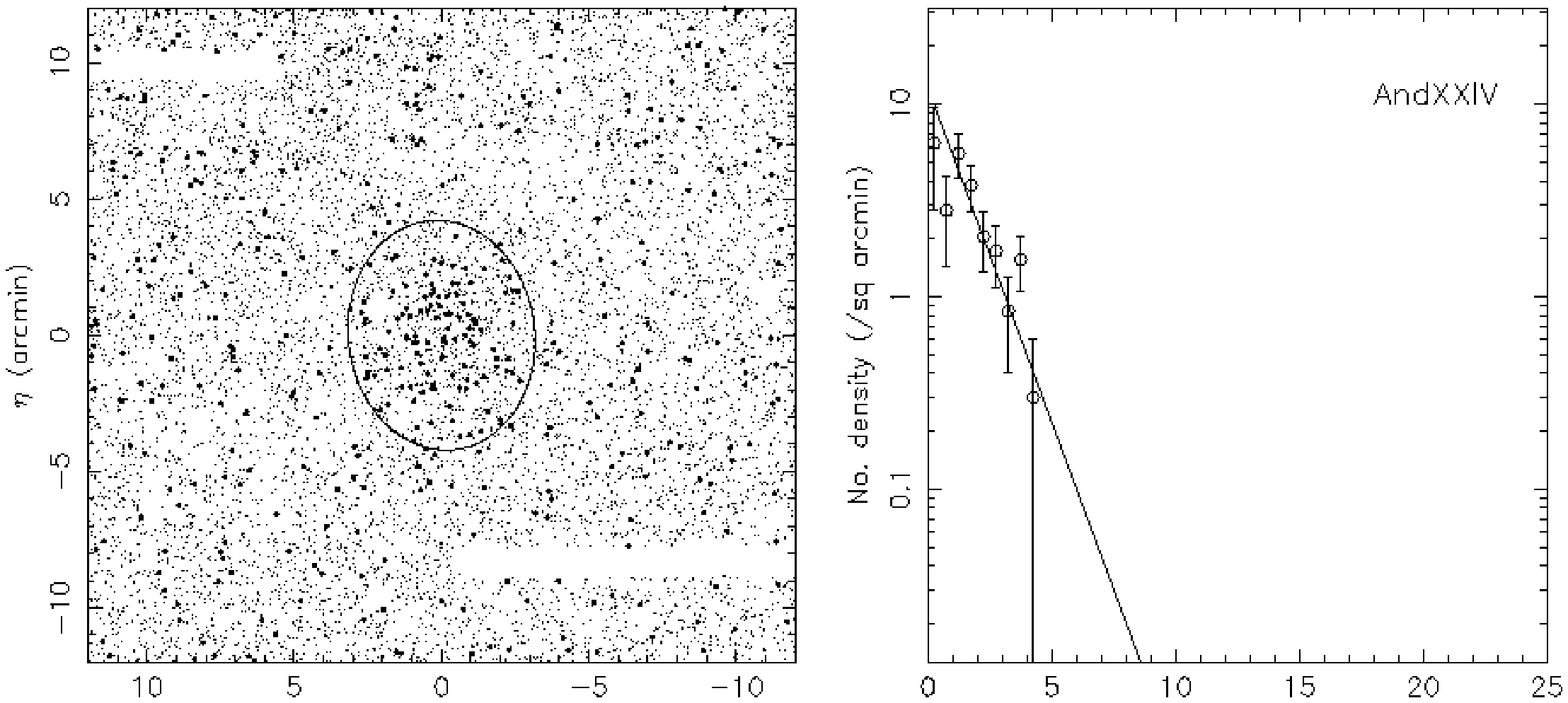}
\vskip0.1cm
\includegraphics[width=0.44\hsize,angle=270]{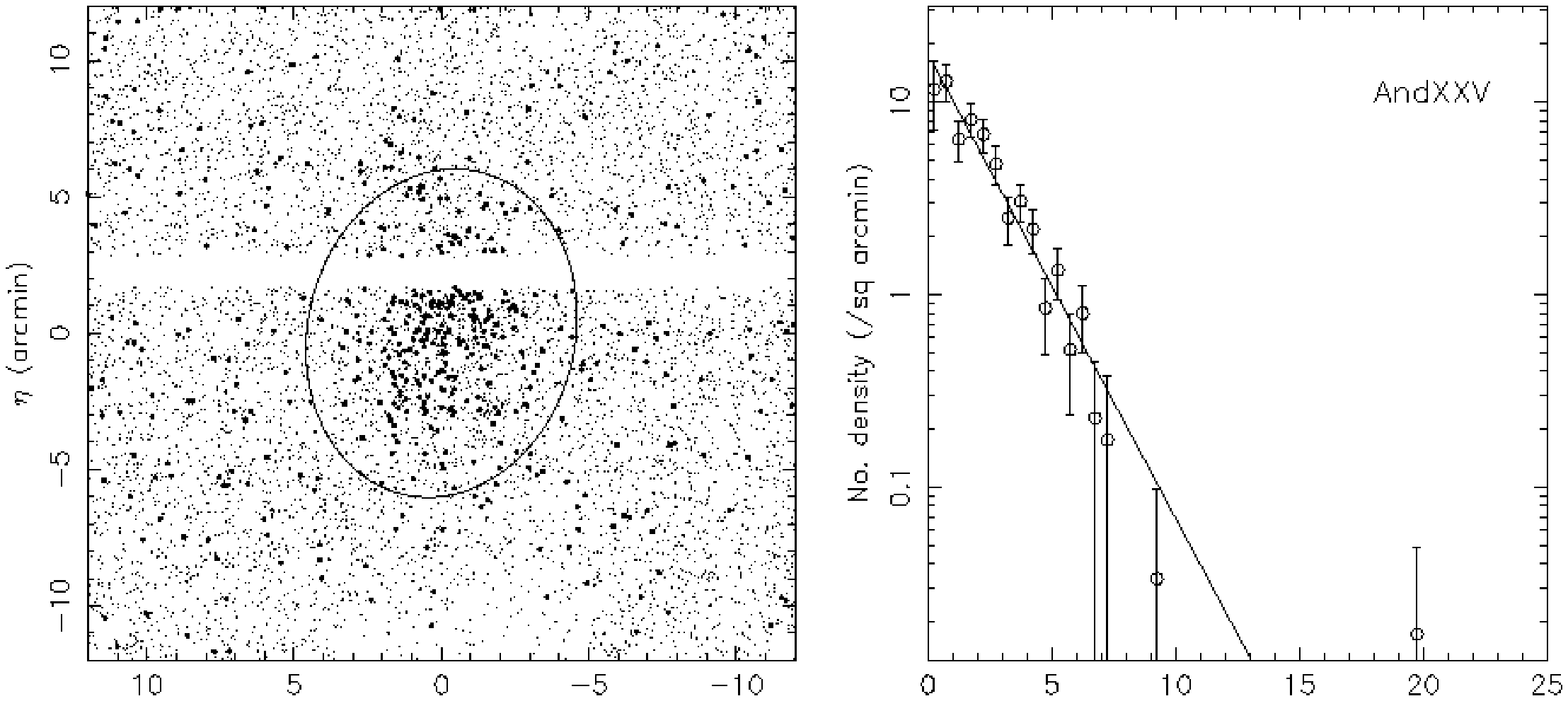}
\vskip0.1cm
\includegraphics[width=0.44\hsize,angle=270]{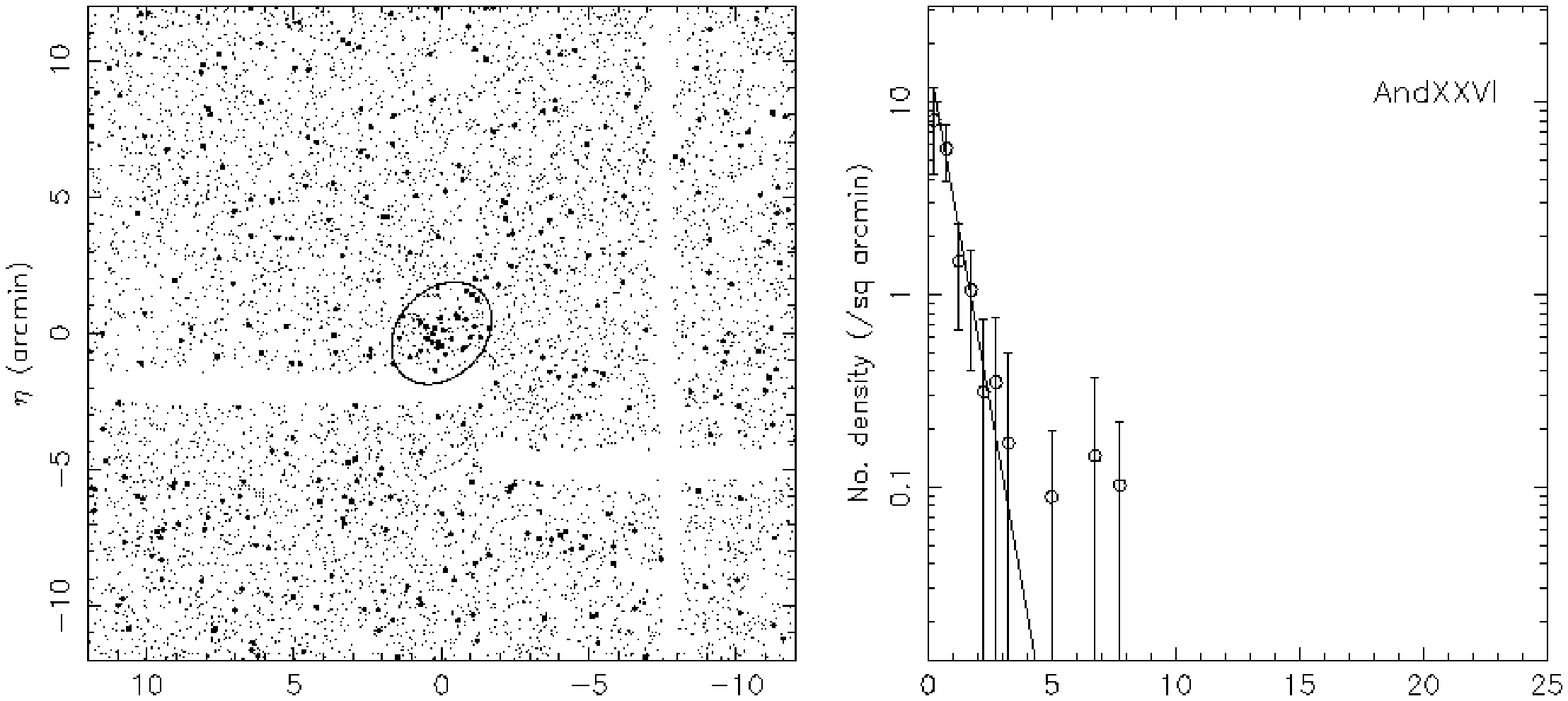}
\vskip0.1cm
\includegraphics[width=0.475\hsize,angle=270]{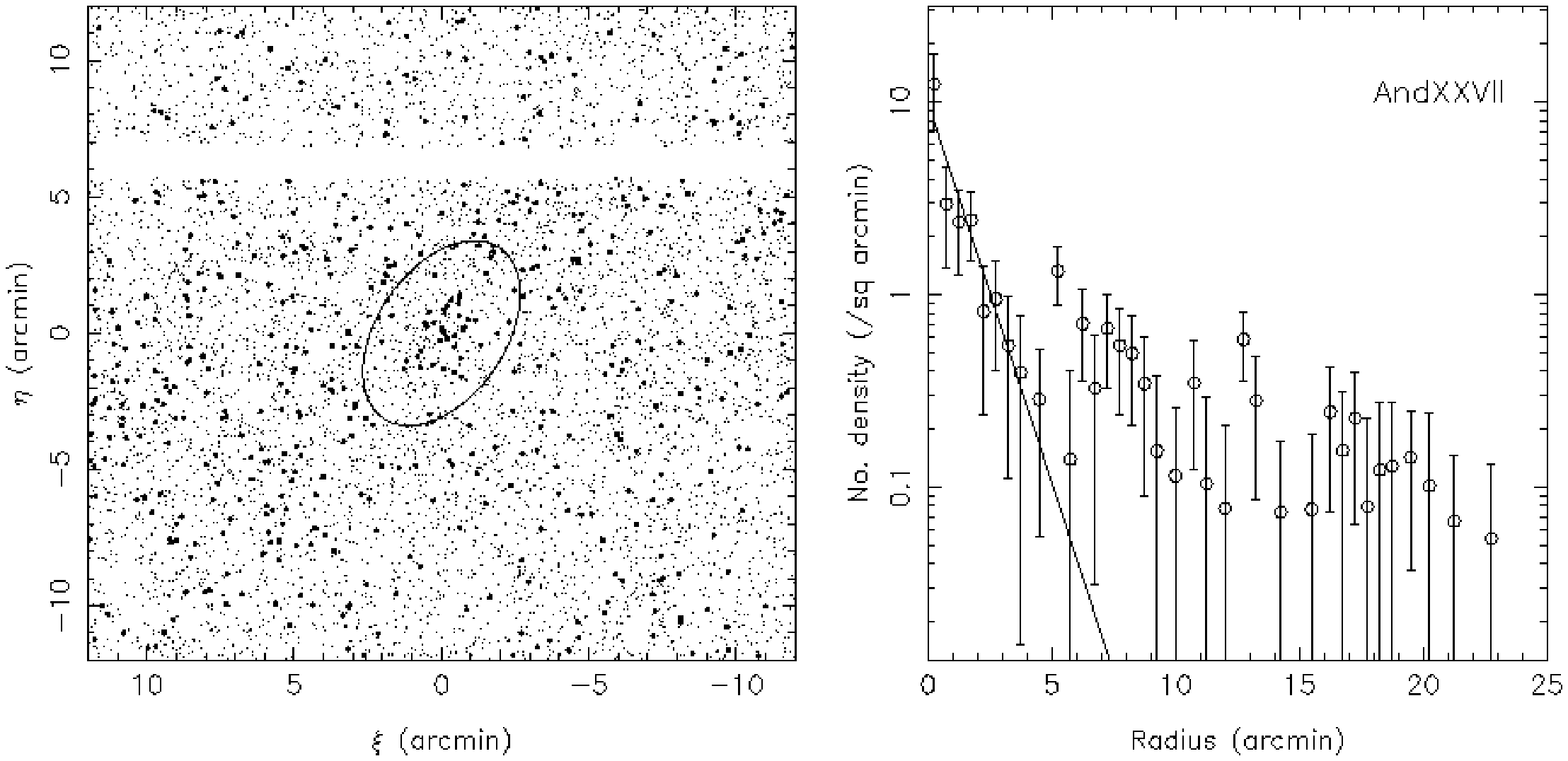}
\caption{\label{fig_spatial}\emph{Left panels:} The spatial
distribution of stellar sources around And~XXIII--XXVII. Small dots
represent all stars in the PAndAS survey and large dots correspond to
stars with colors and magnitudes consistent with metal-poor red giant
branch populations at the distance of M31.  The central ellipse in
each region denotes two half-light radii for each dwarf galaxy using
the structural parameters listed in
Table~\ref{parameters}. \emph{Right panels:} Background-corrected
radial profiles of the dwarfs measured using the average stellar
density within series of fixed elliptical annuli using the parameters
from Table~\ref{parameters}. Allowance is made for incompleteness due
to CCD gaps and the edges of the survey. The error bars account for
Poisson counting statistics and uncertainties in the derived
background level.  The overlaid curves are derived from the tabulated
model parameters (Table~\ref{parameters}) but are not direct fits to
the data points.}
\end{center}
\end{figure}

The five new dwarf galaxies were straightforward to detect as
over-densities in the matched-filter surface density maps of
metal-poor RGB stars.  There are no other obvious candidate satellites
visible in this map or in our equivalent average metallicity and
metal-rich maps.  For simplicity and following convention, we have
named them Andromeda~XXIII - XXVII (And~XXIII - XXVII) after the
constellation they are found in.  As noted earlier the locations of
all previously known dSphs (blue circles) and the five new dwarfs (red
circles) are readily visible in Fig.~\ref{fig_map}.  Moving from East
to West (left to right in the Figure) the new galaxies are And~XXIII,
And~XXIV, And~XXVII, And~XXV and And~XXVI. Their central co-ordinates
are listed in Table~1. A zoom-in view of the distribution of stars
seen in each galaxy is shown in the left-hand panels of
Fig.~\ref{fig_spatial} where the distribution of the candidate
metal-poor RGB stars are highlighted with bolder dots.  The other
points plotted show the rest of the catalogued stellar population in
these fields and serve to highlight where gaps in survey coverage
might have an impact.  The overlaid ellipses are set at twice the
derived half-light radii and also use the ellipticity and position
angle from the likelihood analysis (see
section~\ref{structure}). And~XXIII - XXVII all appear as fairly
obvious over-dense concentrations of metal-poor stars in the localized
spatial distribution maps.

\begin{figure}
\hskip-0cm\includegraphics[width=1.0\hsize,angle=0]{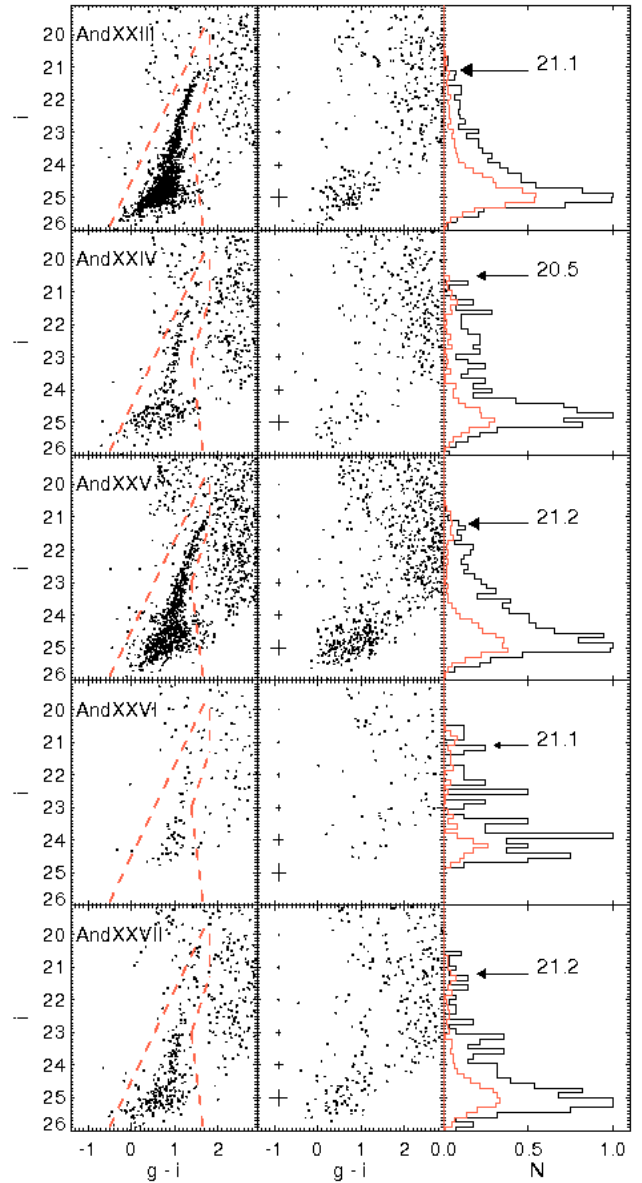}
\caption{\label{fig_cmd}\emph{Left and center panels:} Color-magnitude
diagrams within two half-light radii of And~XXIII--XXVII and of
suitable nearby comparison regions covering the same area after
correcting for gaps in the survey coverage if required.  In all cases
a well-defined RGB is clearly visible as an over-density of stars with
$0.8\lta g-i\lta1.5$ and $i\gta21.0$ that does not have an equivalent
presence in the reference CMD. The {\it rms} photometric errors in i
and g - i as a function of magnitude are displayed in the left-hand
side of the center panel. \emph{Right panels:} The $i$-band luminosity
functions of the central regions of each dwarf (black) and a suitably
scaled much larger neighboring comparison region (red). Only stars
within the dashed polygons were included in the LFs, equivalent
polygons were used to select stars from the reference fields. In most
cases the luminosity functions show a clear change in counts at the
TRGB and the arrows mark the TRGB magnitudes predicted by our distance
estimates (see \S~\ref{distances}).}
\end{figure}

CMDs of stars lying within two half-light radii of the center of each
galaxy (within the ellipses in Fig.~\ref{fig_spatial}; listed in
Table~1) are plotted on the left-hand side of Fig.~\ref{fig_cmd} and
are compared to CMDs of nearby reference regions of the same area
(middle panel of Fig.~\ref{fig_cmd}).  These particular reference
regions were generally defined using an elliptical annulus just beyond
four half-light radii from the center of each dSph.  However, for
And~XXVII this region intersected too much of the enveloping
North-West arc (stream) and an ellipse 5 arcmin to the south and west
was used instead, though even here there is some contamination from
stream stars.  The dwarf galaxies exhibit clear RGB sequences with
$0.8\lta g-i\lta1.5$ and $i\gta21.0$ which are either absent, or
nothing like as apparent, in the reference regions. The narrow color
width of the RGBs, typical of dSph galaxies, suggests that their
stellar populations do not contain a wide range of
metallicities. There is no evidence that any of these galaxies
contains a young main-sequence population. Reference regions are
mainly devoid of RGBs and all contain a `red cloud' of foreground
dwarf stars from the Milky Way (MW) disk at the top right of the CMD
with $g-i\gta2.0$ and $i\lta23.0$. The fainter group of objects
centered at $0.0\lta g-i\lta1.5$ and $i\gta24.0$ and most readily
visible in some of the comparison regions are generally caused by
contamination from mis-classified compact background galaxies.

The right hand panel of Fig.~\ref{fig_cmd} displays the $i$-band
luminosity functions (LFs) of And~XXIII - XXVII (black) computed
within two half-light radii of the centers, and a scaled LF from a
nearby reference field of nine times larger area (red). These large
area reference fields have also been used to measure the distances,
magnitudes and metallicities referred to later in this paper. For
And~XXIII, And~XXIV, And~XXV and And~XXVI the reference regions are
described by elliptical annuli with the same ellipticity and position
angle of the dSph in question (see Table~1), the inner boundary lies
at four half-light radii from the center of the dSph and the outer
boundary is positioned so that the area covered is nine times larger
(after allowing for gaps and edges) than the area used for the dSph.
For And~XXVII a circular reference region was chosen one degree to the
North-East to avoid the extended tidal stream as much as possible.
           
In Fig.~\ref{fig_cmd} only candidate RGB stars, which were selected
from the regions marked with dashed red lines on the dwarf galaxy
CMDs, are included in the LFs. The $i$-band LFs of the corresponding
reference regions, normalized from a much larger area to reduce shot
noise, are unable to match the shape and number over-densities due to
the dSph galaxies.  At bright magnitudes, the LFs peter out and
eventually stop beyond $i \sim 21.0$. This marks the point (TRGB)
where H-shell burning ceases and the He-flash is triggered.  The
I-band luminosity of the TRGB is almost independent of metallicity for
dSph systems such as these \citep{bellazzini01} and provides an
excellent distance indicator for well-populated RGBs.  However, since
most of the new dSphs are relatively sparsely populated and the TRGB
is not always well-defined, we have based our primary distance
estimate on the luminosity of the horizontal branch and indicated the
predicted location of the TRGB with an arrow in Fig.~\ref{fig_cmd}
(see Section~\ref{distances}).

And~XXIII is the largest of the newly discovered dwarf galaxies. Given
its size and stellar density it may seem surprising that it has not
been discovered before but even close inspection of the on-line
digitized photographic plate sky surveys
(e.g. http://archive.stsci.edu/cgi-bin/dss\_form) shows nothing
visible at this location.  However, with central surface brightnesses
of $\Sigma_{V,0} \approx 27.5$ (mag/arcsec$^2$) for all of these new
discoveries (see Section~\ref{magnitude} and Table~1) their
invisibility on previous sky surveys is not unexpected.

The imaging of And~XXV provides a good example of Murphy's Law and
also illustrates the pitfalls of surveys like this with emphasis on
large area coverage and survey speed at the expense of filling all the
gaps between detectors, and also sometimes field edges (see
Fig.~\ref{fig_spatial}).  Much of And~XXV falls in the $\approx$2
arcmin gap between the unbutted edges of two of the rows of 9 CCDs
making up part of the Megacam array.  The small amount of dithering in
the (usual) 3 exposure dither sequence fills the tiny gaps between the
long side of the detectors but does little for the much larger gap
either side of the outer short edges of the middle set of the
2$\times$9 3-edge-buttable detectors.  It is hardly surprising that
gaps and edges make an appearance, even in this small sample of
vignettes, since half of the surveyed region lies within $\approx$5--6
arcmin of either a field boundary or a large gap between rows of
detectors. Fortunately, from a discovery viewpoint, most M31-distance
dSphs are large enough to be found even in they land in the middle of
a gap; while from a census standpoint, the impact of gaps on
completeness is one of several factors that can readily be quantified,
but is outside the scope of this paper.

And~XXIV and And~XXVI have the most sparsely populated RGBs of the
dSphs resulting in much noisier LFs. However, in both cases a
metal-poor RGB sequence can still be clearly discerned, which is
manifestly absent in the companion reference fields. And~XXVI suffers
somewhat for lying in a field with a much shallower limiting $i$-band
and $g$-band depth (by $\approx$0.5\magnitude and 0.25\magnitude
respectively) due to poorer than average seeing) compared to most of
the other fields.

And~XXVII is a unique case among these galaxies as it may be directly
connected to an obvious stellar stream which it seems to form part of.
In Fig.~\ref{fig_map} it can be seen to align with part of the great
North-Western arc of stars, at least in projection. On the CMD, its
stars have the same distribution of colors and magnitudes as stars
selected from the arc implying that they have similar stellar
populations and line-of-sight distances. This suggests the possibility
that And~XXVII is the partially disrupted remnant of a strong tidal
disruption event. Since And~XXVII is embedded in (or superimposed
upon) an extended diffuse stream it was difficult to find a clean
comparison field in the near vicinity and it is likely that the
handful of stars resembling an RGB in the references region
(Fig.~\ref{fig_cmd}) also belong to the stellar stream (and therefore,
potentially, And~XXVII).


\subsection{Structural Properties}
\label{structure}

The structural parameters of And~XXVIII - XXVII were determined by
analyzing the spatial distribution of their resolved metal-poor
candidate stars using a variant of the maximum likelihood technique
developed by \cite{martin08b,martin09} but in this case based on the
Press-Schechter formalism \citep{pressschechter}.  The latter approach
to likelihood problems such as this allows gaps in coverage to be
directly assigned as window functions as an integral part of the
analysis and therefore removes the requirement for artificially
filling the gaps as part of an iterative solution.

Candidate stars used in the analysis are restricted to those used in
the construction of the metal-poor distribution map shown in
Fig~\ref{fig_map}.  As in \cite{martin08b} we use a simple elliptical
exponential profile to describe the over-density of stars in the
galaxy, together with a constant background level.  We differ further
from their procedure by estimating the background level over a much
larger region than used in the likelihood analysis and thereby fixing
this parameter. These background levels are factored into the
uncertainty estimates of the profile parameters but have no effect on
the location parameters. This leaves six remaining parameters to be
estimated.  The central co-ordinates ($\alpha_o$, $\delta_o$),
half-light radius ($r_h$), ellipticity ($\epsilon=1-b/a$ where $a$ and
$b$ are the major and minor axes scale lengths of the system) and
position angle ($\phi$; measured East from North) are derived based on
a simple grid search starting from visually determined initial
estimates. The grid has step sizes of 0.05 in $\epsilon$, $5^{\deg}$
in $\phi$ and $0.05'$ in $r_h$. Providing that the starting central
coordinates lie within the half-light core of the dSph their exact
value has no impact on the derived structural parameters because they
are updated during the solution and the likelihood surface is smoothly
convex around the maximum value. The variation of the likelihood
function over the grid also suffices to define the likelihood surface
around the solution point for use in parameter error estimates (see
for example \citealt{martin08b}). Specifically, the uncertainties in
the profile parameters are quoted as $1\sigma$ errors from direct
analysis of the marginalized likelihood contours from the grid search.

The central over-density $f_o$, is effectively a nuisance parameter
and is readily determined, given the other parameters, by iteratively
solving the non-linear equation $\partial ln\ L / \partial f_o = 0$.
We note that in the Press-Schechter method this is not the same as
using the integral constraint in \cite{martin08b}. The central
co-ordinates were initially chosen by visual inspection of the overall
survey map (which is constructed with an embedded Tangent Plane World
Coordinate System) and then updated by using contoured isopleth maps
of the central regions of each galaxy.  From the viewpoint of their
structural properties defined by $r_h, \epsilon$, $\phi$, the precise
value of the central coordinates ($\alpha_o$, $\delta_o$) are of
secondary importance and we found in practice that we could reduce the
dimensionality of the grid search by updating these directly from the
algebraic solutions of $\partial ln \ L / \partial \Delta\xi = 0$,
$\partial ln \ L / \partial \Delta\eta = 0$ where $\Delta\xi$ and
$\Delta\eta$ are the offsets in standard coordinates (rotated to the
best fitting ellipse coordinate frame) from the current tangent point
($\alpha_o$, $\delta_o$) of the solution (see Appendix for further
details).  After an iteration of the profile parameter grid search,
the central coordinates are updated and the process repeated until a
satisfactory convergence is achieved.

To construct the radial profile for display purposes, the average
density of stars contained within fixed elliptical annuli and at
constant position angle, was evaluated and the profiles were corrected
for foreground and/or background contamination using the (much larger)
normalized reference region.  Due to slight gradients in field
contaminants the error in setting the background level is dominated by
systematics as much as by Poisson noise.  Because of this, as noted
earlier, we decoupled the background determination by fixing the value
in the maximum likelihood estimates of the other parameters and then
directly explored the effects of plausible changes in this background
level by re-running the estimator with a small range of background
values.

The right-hand panels of Fig.~\ref{fig_spatial} show the
background-corrected radial profiles derived from the metal-poor
candidate spatial distributions, where the error bars include
contributions from Poisson errors and the error in determining the
local background level.  The radial distributions shown for each
galaxy were derived from average counts in elliptical annuli using the
derived ellipticities and position angles, with due allowance for gaps
in the survey coverage.  We note that the exponential profile shown
overlaid for each galaxy is not a direct fit to the binned data points
but comes directly from the maximum likelihood analysis outlined
earlier.

The gaps in coverage vary significantly for each galaxy, but even
And~XXV and And~XXVI, the worst affected, are only missing $\sim 10\%$
of their stellar populations and still allow reasonable best fit
parameters to be determined.  And~XXVII presents a different challenge
since here the background model should be more complex than a simple
constant level because of the intersecting tidal stream. The simple
maximum-likelihood method used above failed to converge in this
instance because of the complex stream-like substructure in which
And~XXVII is embedded and we employed an alternative technique to
derive the profile parameters.  To determine $\epsilon$ and $\phi$ we
constructed a contoured isopleth map of the region around And~XXVII
and computed zeroth, first and second moments of the surface density
distribution, at a range of simple thresholds relative to background
(e.g. \citealt{irwin05}). This provided a well-defined $\phi$ but a
relatively poorly defined $\epsilon$ due to the complexity of the
stream-like distribution surrounding the core. Using the average
values of $\epsilon$ and $\phi$ and the weighted center-of-gravity
from the first moments, we then computed the radial elliptical profile
(shown in Fig.~\ref{fig_spatial}) and directly fitted an exponential
profile to the inner 5 arcmin zone.

The derived parameters for all five galaxies are listed in Table~1.
Interestingly these new satellites span the complete range of
parameters found previously in M31 dSphs.  And~XXIII has a
particularly large half-light radius of $r_h = 1.04~\rm{kpc}$. Only
one other currently known dSph satellite companion of M31 has a larger
half-light radius; AndXIX which has $r_h = 1.7~\rm{kpc}$. At the other
extreme, And~XXVI is the smallest of the new galaxies with $r_h =
0.23~\rm{kpc}$ and also among the smallest of the currently known M31
dSph population.

\subsection{Distances}
\label{distances}


\begin{figure}
\hskip-0.3cm\includegraphics[width=1.05\hsize,angle=0]{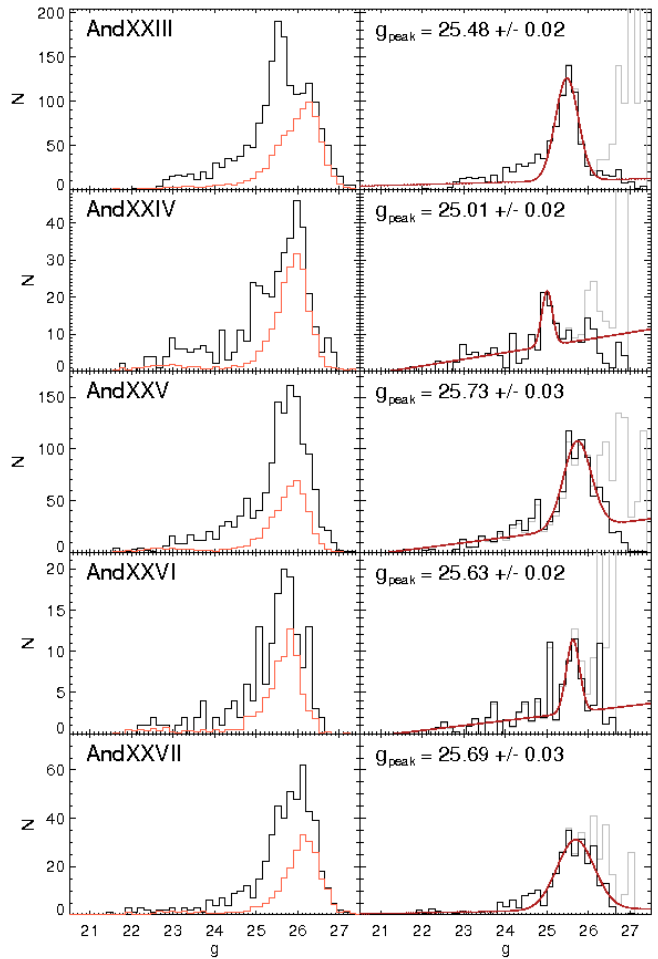}
\caption{\label{fig_glf}\emph{Left panels:} The $g$-band luminosity
functions of the central regions of each dwarf (black) together with
suitably scaled much larger neighboring comparison regions
(red). \emph{Right panels:} Difference luminosity functions (black)
with simple Gaussian plus sloping background fits to the regions
around the horizontal branches (red). An approximate
completeness-corrected difference LF, as described in
Section~\ref{distances}, is plotted in gray. The $g$-band magnitude of
the peak of the fits is indicated together with an estimate of the
$rms$ error in the position of the peak.}
\end{figure}

Most of the newly discovered dwarf galaxies are too poorly populated
to directly determine their distances using traditional TRGB
methods. The absence of sufficient numbers of stars near the tip of
the RGB unsurprisingly leads to unreliable distance estimates (see
Fig.~\ref{fig_cmd}).  Instead we decided to use the extra depth of the
$g$-band for bluer stellar populations to directly measure the
luminosity of the horizontal branch (HB). This feature accounts for
the strong over-densities sloping to fainter magnitudes and bluer
colors at the bottom of the CMDs, e.g. from $ g-i = 1.0, i = 24.5$ to
$ g-i = 0.0, i = 25.5$ on the CMD of And~XXIII (Fig.~\ref{fig_cmd}).
The $i$-band depth cuts through this region precluding direct
isochrone fitting but we can still exploit the better depth of the
$g$-band directly by analyzing the $g$-band luminosity function
instead.

As a benchmark, the HB magnitude of M31 is $g_0 = 25.2$, which in the
typical extinction for this region translates to an apparent magnitude
of $g = 25.4 - 25.5$.  Since most of the $g$-band data we have extends
down to at least $g = 26.5$ mag (and to between $g = 25.5-26.0$ with
S/N = 10), we have a good chance of resolving the HBs of the new
dSphs. However, not surprisingly, the comparison regions in the CMDs
show that there is a notable component of contamination from
unresolved background galaxies which necessitates a careful treatment
of the background correction.

The observed $g$-band LFs of And~XXIII - XXVII are shown in the
left-hand panel of Fig.~\ref{fig_glf} (black histograms) and those of
the same appropriately scaled background reference fields (see
Section~\ref{desc}) used for the $i$-band LFs are shown in red. Since
the $g$-band data reaches significantly deeper than the $i$-band for
such stars, only a detection in the $g$-band is required and the
number of stars contributing to these LFs is much larger than for the
equivalent $i$-band LFs. The right-hand panel of Fig.~\ref{fig_glf}
shows the $g$-band LF of each galaxy with the background-subtracted
(black histogram) and a best fit Gaussian (including a term to account
for the sloping background of the LF) to the HB over-plotted in
red. The peak of the fits and corresponding rms errors for each galaxy
are shown in Fig.~\ref{fig_glf}. The peak in the $g$-band luminosity
corresponding to the HB varies from g = 25.0 mag to $g = 25.7$
mag. The rms error in the $g$-band magnitude over this range is
$\approx$0.1 mag and the putative HB fits are generally well-defined.

To ascertain the sensitivity of the HB estimates to completeness
effects we have generated approximate completeness-corrected
difference LFs using a generic form for the completeness as a function
of point-source S/N. Full completeness tests are outside the scope of
this paper but we note that in comparable datasets the completeness
level is better than $90\%$ for $10\sigma$ fluxes and typically drops
to $\sim50\%$ by $5\sigma$.  For the galaxies presented here, a S/N of
$5\sigma$ occurs for $g$-band magnitudes of 26.0 - 26.3, where the
clearest impact of this an apparent steep fall-off in the LF of the
comparison regions. The approximate completeness-corrected difference
LFs are plotted as gray histograms in the right-hand panels of
Fig.~\ref{fig_glf}. They show that the HB is well resolved for
And~XXIII, And~XXIV, And~XXV and And~XXVI. For these galaxies the
number of stars reaches a clear peak (highlighted by the Gaussian fit)
around the HB and drops down again before beginning to rise as the RGB
progresses to fainter magnitudes.  The best-fit Gaussians to the HB
peaks of the completeness-corrected LFs differed from the original
fits by $\leq 0.02$~mag, corresponding to a maximum deviation in the
final distances of 9~kpc, which is negligible compared to other
uncertainties.

In the case of And~XXVII the completeness-corrected LF distorts the HB
peak, although this is exacerbated by the difficulty of finding a
clean nearby comparison region. Consequently here we define a lower
limit only for this galaxy. The peak of stars at g = 25.5~mag on the
bottom right hand panel of Fig.~\ref{fig_glf} marks the brightest
possible magnitude of the HB and we use this to determine the lower
limit on the distance.

To calculate the distance modulus we must first estimate a standard
absolute magnitude, M$_{\rm g}$ for the HB populations typical of
these dSphs. Empirical models which estimate the closely related
M$_{\rm V}$ of the HB (e.g. \citealt{gratton98}) have been
predominantly determined from globular clusters and are shown to have
a weak metallicity dependence (gradient). At the range of
metallicities expected here, $-2.5 < \FeH < -1.5$, the variation is
less than $\pm0.1$\magnitude which suggests a simple calibration based
on similar dSphs which have independently determined reliable
distances from TRGB measures and/or HB isochrone fits from HST
data. To determine the value of the constant we additionally measured
the HB $g$-band luminosity of four relatively bright and well-studied
dSphs (And~I, And~II, And~III and And~XVI) which are covered by the
PAndAS survey. The distance moduli of these well-populated dSphs,
which have been determined using the TRGB method in previous papers,
are $24.43 \pm 0.07$ \citep{mcconnachie04}, $24.05 \pm 0.06$
\citep{mcconnachie04}, $24.38 \pm 0.06$ \citep{dacosta02} and $23.60
\pm 0.20$ \citep{ibata07}. Since these galaxies contain hundreds to
thousands of RGB stars the {\it rms} errors in these measurements are
small. Another reason for choosing these particular galaxies is
because they tend to lie on the MW side of M31 and hence have
clearly-defined HBs in their $g$-band LFs.  The formal {\it rms} error
in locating their HB magnitudes is negligible ($\sim 0.02-0.03$~mag)
compared to the systematic errors due to the precise mix of their
stellar populations

By comparing to the known distance moduli of And~I, And~II, And~III
and And~XV we find values for the absolute magnitude M$_{\rm g}$ of
0.89, 0.89, 0.79 and 0.67 respectively and have therefore adopted a
constant of M$_{\rm g} = 0.8\pm0.1$ in the CFHT MegaCam AB magnitude
system. These values are consistent, within the errors, with those
derived for red HB stars by \cite{chen09} using SDSS data for eight
globular clusters with similar metallicities.  The distances estimated
by this method are dominated by the uncertainty of M$_{\rm g}$, which
in turn reflects uncertainties in the mix of blue-HB, red-HB, and red
clump (RC) populations in these galaxies.  Metallicity variation is
most likely a secondary aspect given that all of the new objects have
similar overall metallicities, [Fe/H] = -1.7 to -1.9, (see
section~\ref{metallicities}) to the calibrating objects ([Fe/H] =
-1.5, -1.5, -1.9 and -1.7 from \citealt{mcconnachie04, dacosta02,
ibata07} respectively).  Age effects will also influence the
properties of the HB and RC \eg \cite{girardi01}, however, given their
appearance on the CMDs (Fig~\ref{fig_cmd}) it is highly unlikely that
the newly discovered dSphs contain anything other than old, metal-poor
stellar populations. To further refine these distance estimates deeper
targeted imaging of sufficient depth to allow isochrone fitting to
beyond the HB magnitude is required.

The distance moduli of the five new galaxies, after due allowance for
line-of-sight reddening, are listed in Table~1. The errors in the
distance moduli were calculated by propagating the error in M$_{\rm
g}$, the error in the measured location of the HB (see
Fig~\ref{fig_glf}) and the $\pm3\%$ uncertainty in the calibration of
the photometry. And~XXIV lies the closest to the MW at $600 \pm
33~\rm{kpc}$, And~XXIII and And~XXVI lie at roughly the same
line-of-sight distance as M31, $767 \pm 44~\rm{kpc}$ and $762 \pm
42~\rm{kpc}$ respectively, while And~XXV lies furthest from us, beyond
M31, at $812 \pm 46~\rm{kpc}$. The difficulty in measuring the HB
magnitude of And~XXVII means we can only assign a lower limit to its
distance of $\geq 757 \pm 45~\rm{kpc}$.

As an independent consistency check we have calculated the predicted
magnitude of the TRGB in the $i$-band using our HB-based distance
moduli and indicated them with an arrow in Fig.~\ref{fig_cmd}. The
estimated locations of the TRGBs of And~XXIII - XXV agree well with
the drop off in star counts at the bright end of their LFs. For
And~XXVI and And~XXVII the case is less clear due to the small numbers
of stars, increasing the contamination from foreground stars from the
overlapping `red cloud' region. To reliably disentangle dSph members
from foreground stars in this part of the CMD requires spectroscopic
followup (\eg \citealt{letarte09}).

Further supporting evidence for the distances we have calculated is
provided in the right-hand panels of Fig~\ref{fig_mdf}. Here
theoretical isochrones have been shifted to the derived distances and
compared to the CMDs of the galaxies. In each case the isochrones
agree favourably with the distribution of the RGB stars.

\subsection{Magnitudes}
\label{magnitude}

The absolute magnitudes ($M_V$) were computed by comparing the LFs of
And~XXIII - XXVII to the LFs of And~I, And~II and And~III.  These
latter dSphs are also part of the PAndAS survey and cover a similar
range of metallicities and (presumably) mix of stellar populations.
Absolute magnitudes of $M_V~\rm{(AndI)} = -11.8 \pm 0.1$,
$M_V~\rm{(AndII)} = -12.6 \pm 0.2$ and $M_V~\rm{(AndIII)} = -10.2 \pm
0.3$ were taken from \cite{mcconnachie04} and \cite{mcconnachie06a}.
For each dSph background-subtracted $g$-band luminosity functions were
constructed using stars within loci such as those marked on the CMDs
in Fig~\ref{fig_cmd}.  All LFs are corrected for CCD-gaps and include
only stars within two half-light radii of the center of each galaxy
using the structural parameters listed in Table~1 for the new dSphs
and in \cite{mcconnachie06a} for the reference objects.  Limiting the
selection to two half-light radii includes roughly 85\% of the flux
and lessens the sensitivity to bright interlopers, but still leaves a
shot-noise dominated estimate at the bright end.

We then select the portion of the LFs containing all stars above the
$\approx$ 90$\%$ completeness level ($g < 25.5~\rm{mag}$) and below an
upper flux limit ($g > 22.5~\rm{mag}$) for all eight dSphs.  Assuming
the LFs are similar, the maximum likelihood scale factors that yield
the best match between the new dwarfs and the reference set (And~I,
And~II and And~III), defines the difference in magnitude between the
galaxies.

The variation in computed absolute magnitudes for each reference dSph
gives some idea on the uncertainty introduced by assuming that the LFs
of all of these galaxies are the same, when they may have different
underlying stellar populations. Similarly shifting the LFs by the
maximum/minimum distance moduli allowed by our measurements in
Section~\ref{distances} gives a handle on the uncertainties introduced
by the distances.  As an example, the absolute magnitudes of And~XXVI
computed in this way range from -6.6 to -7.5~mag giving $M_V = 7.1 \pm
0.5$. The final absolute magnitudes we quote are the mean of each set
of three results and we estimate the error to be $\pm$0.5\magnitude
based on the average scatter in the computed sets of absolute
magnitudes.  As a consistency check we repeated the estimates using
the $i$-band luminosity functions computed in a similar way but
covering the magnitude range $21.0 < i < 24.5$ and found results
consistent to within $\pm$0.1~mag in all cases.

The derived absolute magnitudes of the new dwarfs are listed in
Table~1 and range from $M_V = -10.2 \pm 0.5$ for And~XXIII to $M_V =
-7.1 \pm 0.5$ for And~XXVI. These absolute magnitudes were then used
in conjunction with the derived structural parameters to directly
estimate the central surface brightnesses, $\Sigma_{V,0}$, of the
galaxies, also listed in Table~1. All of the new dSphs have similar
central surface brightnesses ranging from $\Sigma_{V,0} =
28.0~\rm{mag/arcsec}^2$ for And~XXIII to $\Sigma_{V,0} =
27.1~\rm{mag/arcsec}^2$ for And~XXV.

\subsection{Metallicities}
\label{metallicities}

\begin{figure}
\hskip-0.2cm\includegraphics[width=1.05\hsize,angle=0]{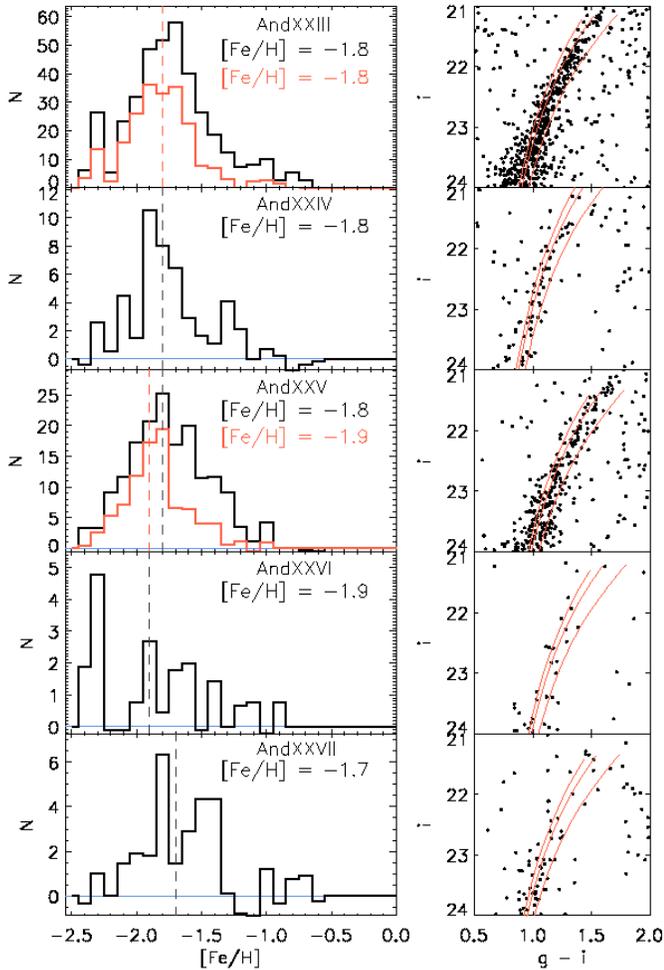}
\caption{\label{fig_mdf}\emph{Left panels:} Metallicity distribution
functions (MDFs) for And~XXIII-XXVII derived assuming a fixed age of
12~Gyr an alpha-element enhancement of [$\alpha$/Fe] $=$ 0.0 and using
the Dartmouth isochrone set \citep{dotter08}. Only stars within two
half-light radii, with $21.0 < i < 24.0$ and colors bluer than the
[Fe/H] = -0.5 dex isochrone were considered.  A correction for
background contamination was made using a suitably normalized (larger)
nearby comparison region and the zero-point is marked with a blue
line. The median metallicity of each MDF is indicated by a dashed gray
line. All of the dSphs have similar median metallicities ranging from
[Fe/H] = -1.7 to [Fe/H] = -1.9 $\pm 0.2$. For the two most populated
galaxies, And~XXIII and And~XXV, we have also derived the MDF of stars
with $i \leq 23.0$ (red histograms) as a check on the impact of
photometric errors.  \emph{Right panels:} show a zoom in section of
the RGBs overlaid with \citet{dotter08} isochrones with age = 12~Gyrs
and metallicities of [Fe/H] = -2.5, -2.0 and -1.5 which have been
shifted to the distance modulus calculated for each galaxy. The
position of the isochrones is in very good agreement with the
magnitude range and shape of the galaxy RGBs.}
\end{figure}

Metallicity Distribution Functions (MDFs) of And~XXIII - XXVII are
presented in the left-hand panel of Fig.~\ref{fig_mdf}. We have used
the \cite{dotter08} isochrones in the CFHT/MegaCam g, i system
(private comm.) to construct the MDFs so that no filter
transformations were required. Specifically, the data are compared to
a grid of isochrones with age = 12~Gyrs, $[\alpha/$Fe] = 0.0 and $-2.5
\leq$ [Fe/H] $\leq -0.5$, spaced in 0.1~dex steps, which have been
reddened to match the data and shifted to the distance modulus we
determined for each galaxy (the E(B-V) and (m - M)$_0$ values used are
listed in Table~1). The right-hand panel shows how the resulting
isochrones for metallicities of [Fe/H] = -2.5, -2.0 and -1.5 dex
compare to the RGB sequences of And~XXIII - XXVII. The -2.5 dex and
-1.5 dex isochrones bound the RGB and match its curvature and
brightness well in all cases, providing further support for our
distance estimates. Stars within two half-light radii of the central
co-ordinates of each galaxy were included in the MDFs if they
satisfied $21.0 < i <24.0$ and were bounded by the -2.5~dex and
-0.5~dex isochrones. Each star is assigned the metallicity of the
nearest isochrone and a histogram of the resulting distribution of
metallicities is constructed. The MDFs have been background subtracted
using a suitably scaled (larger area) reference field satisfying the
same conditions.

We have determined the median metallicity of each galaxy and marked it
with a gray dashed line in Fig.~\ref{fig_mdf}. Within the errors, all
of the galaxies have the same median metallicity ranging from [Fe/H] =
$-1.7 \pm 0.2$~dex to [Fe/H] = $-1.9 \pm 0.2$~dex.  We note that the
errors due to uncertainties in distance modulus have little impact on
the upper part of the RGB for such low metallicity systems.
Re-deriving the metallicities of all five dSphs when the isochrones
are shifted by $\pm$0.1~mag in distance modulus results in change of
median [Fe/H] by $\mp$0.1~dex.  Altering the fiducial value of
$[\alpha/$Fe] from 0.0 to 0.4 would systematically change their
average [Fe/H] by -0.2 dex, while a $\pm3\%$ photometric error in g -
i would result in a shift of $\pm$0.1~dex.

The MDFs of And~XXIII and And~XXV, the two most populated galaxies,
have very broad distributions (FWHM $\sim$0.5 - 0.8~dex).  We caution
that this should not be interpreted too literally as a large
metallicity spread. Although there is some evidence at brighter
magnitudes for a significant spread in the locus of the RGB for
And~XXIII and And~XXV, for the more metal-poor isochrones, photometric
errors, particularly at the faint end, contribute significantly to the
spread in the derived MDF and deeper targeted observations are needed
to further investigate this.  The red histograms in Fig.~\ref{fig_mdf}
are the MDFs derived for And~XXIII and And~XXV when only stars
brighter than i = 23.0 are considered. Both distributions,
particularly that of And~XXV, are somewhat narrower than before, but
although photometric errors at fainter magnitudes contribute to
broadening the MDF, they have little effect on the derived median
[Fe/H] values. The sparseness of the RGB for the fainter dSphs makes
it difficult to assess any spread in the locus though we note that
And~XXIV has a particularly narrow spread of derived metallicities
around its median value (FWHM = 0.3~dex), and is perhaps the galaxy
with the clearest sign of having a simple stellar population. The MDFs
of And~XXVI and And~XXVII are too noisy to make comparable inferences
from their distributions.


\begin{deluxetable*}{l c c c c c c}
   \tablewidth{18cm} \tablecaption{\label{parameters}
     Derived properties of the satellites} \tablehead{
     \colhead{Parameter} & \colhead{And~XXIII} & \colhead{And~XXIV} & 
     \colhead{And~XXV} & \colhead{And~XXVI} & \colhead{And~XXVII} }
     \startdata

$\alpha$ (J2000) & $01^{\rm h}29^{\rm m}21.8^{\rm s}\pm 0.5^{\rm s}$ &
                   $01^{\rm h}18^{\rm m}30.0^{\rm s}\pm 0.5^{\rm s}$ &  
                   $00^{\rm h}30^{\rm m}08.9^{\rm s}\pm 0.5^{\rm s}$ &
                   $00^{\rm h}23^{\rm m}45.6^{\rm s}\pm 0.5^{\rm s}$ &  
                   $00^{\rm h}37^{\rm m}27.1^{\rm s}\pm 0.5^{\rm s}$ \\
$\delta$ (J2000) & $38\deg43'08''\pm10''$ & $46\deg21'58'' \pm10''$ & 
                   $46\deg51'07''\pm10''$ & $47\deg54'58'' \pm10''$ &
                   $45\deg23'13''\pm10''$ \\
$(l,b)$ ($\deg$) & $(131.0,-23.6)$ & $(127.8,-16.3)$ &
                   $(119.2,-15.9)$ & $(118.1,-14.7)$ & $(120.4,-17.4)$ \\
$E(B-V)^\mathrm{(a)}$ & 0.066 & 0.083 & 0.101 & 0.110 & 0.080 \\
$(m-M)_0^\mathrm{(b)}$ & $24.43\pm0.13$ & $23.89\pm0.12$ & $24.55\pm0.12$ & 
            $24.41\pm0.12$ & $24.59\pm0.12$  \\
$D$ (kpc) & $767\pm44$ & $600\pm33$ & $812\pm46$ & $762\pm42$ & $\geq757\pm45$ \\
$r_\mathrm{M31}$ (kpc) & $\sim126\pm44$ & $\sim197\pm33$ & $\sim97\pm47$ & 
                       $\sim101\pm42$ & $\sim86\pm48$  \\
$\rm{[Fe/H]}^\mathrm{(c)}$ & $-1.8\pm0.2$ & $-1.8\pm0.2$ & $-1.8\pm0.2$ & $-1.9\pm0.2$ & $-1.7\pm0.2$  \\
$r_h \rm{(arcmin)}$ & $4.6\pm0.2$ & $2.1\pm0.1$ &  $3.0\pm0.2$ &
                 $1.0\pm0.1$ &  $1.8\pm0.3$:  \\
$r_h (pc)$ & $1035\pm50$ & $378\pm20$ & $732\pm60$ &
                   $230\pm20$ & $455\pm80$: \\
$\phi$ (N to E) ($\deg$) & $138\pm5$ & $5\pm10$ & $170\pm10$ &  
                              $145\pm10$ & $150\pm10$: \\
$\epsilon=1-b/a$ & $0.40\pm0.05$ & $0.25\pm0.05$ & $0.25\pm0.05$ &
                   $0.25\pm0.05$ & $0.4\pm0.2$:  \\
$M_V$ & $-10.2\pm0.5$&$-7.6\pm0.5$&$-9.7\pm0.5$ & $-7.1\pm0.5$ & $-7.9\pm0.5$ \\
$\Sigma_{V,0}$ (mag/arcsec$^2$) & $28.0\pm0.5$ & $27.8\pm0.5$ &
          $27.1\pm0.5$ & $27.4\pm0.5$ &  $27.6\pm0.5$  \\

     \enddata

\footnote{\textbf{Notes.}The errors quoted for the profile parameters
derived in \S~\ref{structure} ($\alpha, \delta, r_h, \phi$ and
$\epsilon$) are the $1\sigma$ errors from the marginalized likelihood
contours of the grid search. Structural parameters derived for
And~XXVII are marked ':' to reflect the additional uncertainties
caused by the presence of the intersection stellar
stream. $^\mathrm{(a)} E(B-V)$ interpolated from the Schlegel et
al. (1998) dust maps. $^\mathrm{(b)}$ Due to the sparsity of the upper
RGB, distance moduli are calculated from the magnitude of the
horizontal branch from the deeper $g$-band data. $^\mathrm{(c)}$ The
median metallicity of the metallicity distribution functions from
Fig.~\ref{fig_mdf}.}
\end{deluxetable*}

\section{Discussion}

\subsection{Structural properties}


\begin{figure}
\begin{center}
\vskip 0.2cm
\hskip-0.2cm\includegraphics[width=1.0\hsize,angle=0]{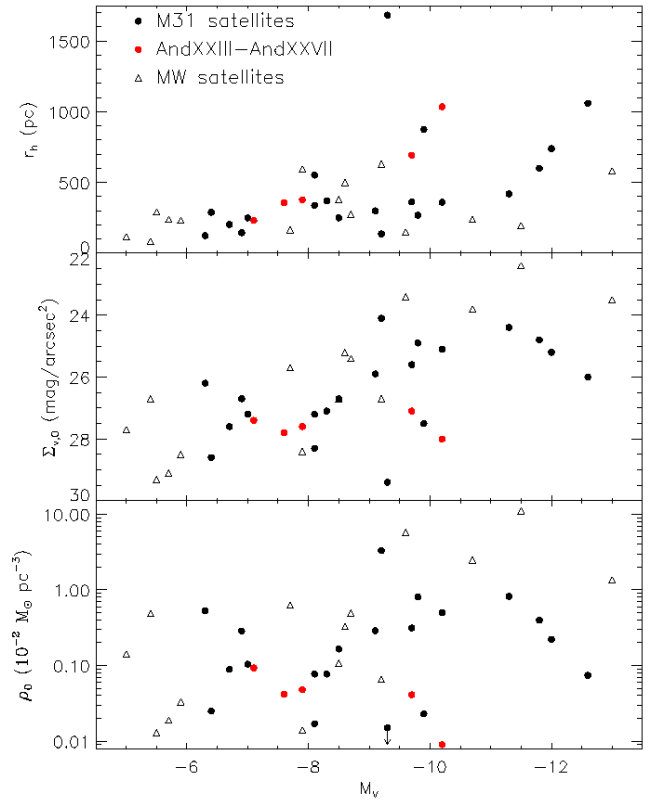}
\caption{\label{fig_relations}\emph{Top panel:} The variation of
half-light radius ($\rm{r}_h$) with absolute magnitude ($\rm{M}_V$)
for the five newly discovered galaxies presented in this paper (red
circles), other M31 satellites (black circles) and MW satellites
(triangles). The source of the data is referred to in the
text. \emph{Middle panel:} The variation of central surface brightness
($\Sigma_{V,0}$) as a function of $\rm{M}_V$. Values of $\Sigma_{V,0}$
have been taken from the literature where available, otherwise they
have been estimated using $\rm{r}_h$ and $\rm{M}_V$ as described in
\S~5.1. \emph{Bottom panel:} The variation of inferred central
luminosity density $\rho_o$ as a function of $\rm{M}_V$. And~XIX has a
value much lower than the plot boundary as indicated by the downwards
arrow.  Note both the factor of 1000 variation in central luminosity
(and implied baryon) density over the whole population and the factor
10 difference in this quantity between the brighter ($M_V < -9.0$) M31
dSphs and their MW counterparts.}
\end{center}
\end{figure}

The five new dSphs have characteristics typical of the range of
properties of the previously known M31 dSph population.  The top panel
of Fig.~\ref{fig_relations} shows their location in the (semi-major
axis) half-light radius --v-- absolute magnitude plane together with
the location of the other M31 dSph satellites and equivalent MW
satellites satisfying $\rm{M}_V \leq -4.5$.  The values for the MW
satellites Fornax, Leo~I, Sculptor, Leo~II and Sextans were taken from
\cite{irwin95} while the parameters for the rest of the MW satellites
were taken from \cite{okamoto08} and Okamoto (PhD thesis) and include
the MW ultra faint dwarfs (UFDs) Boo I, Hercules, CVn I, CVn II, Leo
IV, Leo T and Uma I.  The equivalent data for the previously known M31
satellites is from \cite{mcconnachie06a, martin06, ibata07,
mcconnachie08, martin09}.

As expected, there is a general correlation between the size of the
galaxy and its magnitude, though we caution that fainter than
$\rm{M}_V \approx -8$ observational selection effects may be limiting
the detectability of larger half-light radii systems.  The MW
satellite population extends to fainter magnitudes than M31 satellites
because, at the distance of M31, even relatively compact dwarf
galaxies fainter than $\rm{M}_V \approx -6$ are beyond our current
detection limits.  Although this rules out any direct constraints on
putative UFDs around M31 it is interesting that the number of M31
satellites per absolute magnitude interval shows no sign of tailing
off towards the faint end.  This suggests that an equivalent
population of UFDs are waiting to be discovered.

Following their analysis of the structural properties of And~I--VII, 
\cite{mcconnachie06a} noted that the dwarf satellites of M31
tended to have larger radii than MW dwarf satellites and were
typically twice as big. Subsequently, 18 new M31 dSphs (And~XI -
And~XIII \citep{martin06}, And~XIV \citep{majewski07}, And~XV and
And~XVI \citep{ibata07}, And~XVII \citep{irwin08}, And~XVIII - And~XX
\citep{mcconnachie08}, And~XXI and And~XXII \citep{martin09}) and 10
new MW dSphs, including six with $\rm{M}_V \leq -4.5$, have been
discovered: CVn~I \citep{zucker06a}; Boo I \citep{belokurov06};
Hercules, CVn~II and Leo~IV \citep{belokurov07}; and Leo~T
\citep{irwin07}. Recent work by \cite{kalirai10} found that there is
significant overlap between the M31 and MW satellite populations at
the low luminosity end ($L \lesssim 10^6~L_{\sun}$).

Fig.~\ref{fig_relations} includes data from the new M31 and MW
satellite galaxies more luminous than $\rm{M}_V = -4.5$ in addition to
the previously known satellites and the newly discovered galaxies
presented for the first time in this paper, And~XXIII--XXVII.  When
the new data are included there is no clear difference in the
$\rm{M}_V - \rm{r}_h$ relation between MW and M31 satellites for
absolute magnitudes fainter than $\rm{M}_V \approx -9$.  However, for
the subset of dSph galaxies brighter than $\rm{M}_V \approx -9$ it
remains the case that the M31 dSph satellites are generally twice as
extended as their MW counterparts, even though they are found at a
similar range of distances ($\approx100 - 300\kpc$) from their host
galaxy.  Of the new dSphs, the brighter pair, And~XXIII and And~XXV,
continue this trend, whilst the three with $M_V > -9$ (And~XXIV,
And~XXVI and And~XXVII) have half-light radii typical of both M31 and
MW dSphs in this magnitude range.  It is unlikely that this is a
completeness issue since MW satellite galaxies brighter than $\rm{M}_V
= -9$ and with $\rm{r}_h \geq 600$~pc should be easy to find if they
lie within the expected distance range of MW satellites (\eg
\citealt{koposov08}).

As an alternative way of looking at these distributions, the variation
of central surface brightness, $\Sigma_{V,0}$, as a function of
half-light radius is shown in the middle panel of
Fig.~\ref{fig_relations}.  Fainter than $\rm{M}_V \approx -9$, MW and
M31 satellites show a similar range of central surface brightness.
However, for those brighter than $\rm{M}_V \approx -9$, M31 satellites
have significantly fainter central surface brightness than their MW
counterparts by an average of $\sim$2~mag/arcsec$^2$.

Finally, in the third panel of Fig.~\ref{fig_relations} we make the
plausible assumption that, to first order, we can deconvolve the
projected surface density distribution by approximating it as a
Plummer law \citep{plummer11} with the geometric mean half-light
radius deduced from the parameters in Table~1.  This allows us to
analytically infer the equivalent central luminosity density and also,
by effectively assuming a constant baryonic mass:to:light ratio, gain
a measure of the variation of the central baryon density.

In M31 it appears that the brighter (classical) dSphs are not only
twice as extended as their MW counterparts but also correspondingly
have central stellar densities a factor of ten smaller.  This is a
significant difference and suggests that the local environment around
these two L$_*$ galaxies has exerted a strong influence on the
formation and evolution of these dSph systems.

In a study attempting to relate the observed properties of Local Group
dSphs to their dark matter content, \cite{penarrubia08} suggested
that, under the assumption that M31 and MW satellites have similar
dark matter halos, the systematic difference in their sizes should be
complemented by a systematic difference in their
kinematics. Specifically, they predicted that M31 satellites should
have a velocity dispersion $\sim50\% - 100\%$ larger than the
corresponding MW satellites. However, a recent kinematic survey of
And~I, And~II, And~III, And~VII, And~X and And~XIV by \cite{kalirai10}
and similar work by \cite{collins10} has shown that this prediction is
not borne out by observation. M31 satellites with measured velocity
dispersions range from And~X at $\rm{M}_V = -8.1$ and $\sigma_v = 3.9
\pm 1.2 \rm{km s}^{-1}$ to And~I at $\rm{M}_V = -11.8$ and $\sigma_v =
10.6 \pm 1.1 \rm{km s}^{-1}$ \citep{kalirai10}. Meanwhile, MW
satellites covering the same absolute magnitude interval have
$\sigma_v = 9.5 \pm 2.0 \rm{km s}^{-1}$ (Draco at $\rm{M}_V = -8.3$)
to $\sigma_v = 10.5 \pm 2.0 \rm{km s}^{-1}$ (Fornax at $\rm{M}_V =
-13.0$) according to \cite{mateo98}. This suggests, as also noted by
\cite{penarrubia08}, that the nature of the dark matter halos
themselves may be responsible for the observed differences between the
two satellite systems.

\subsection{Satellite spatial distribution}


\begin{figure*}
\begin{center}
\includegraphics[width=0.475\hsize,angle=270]{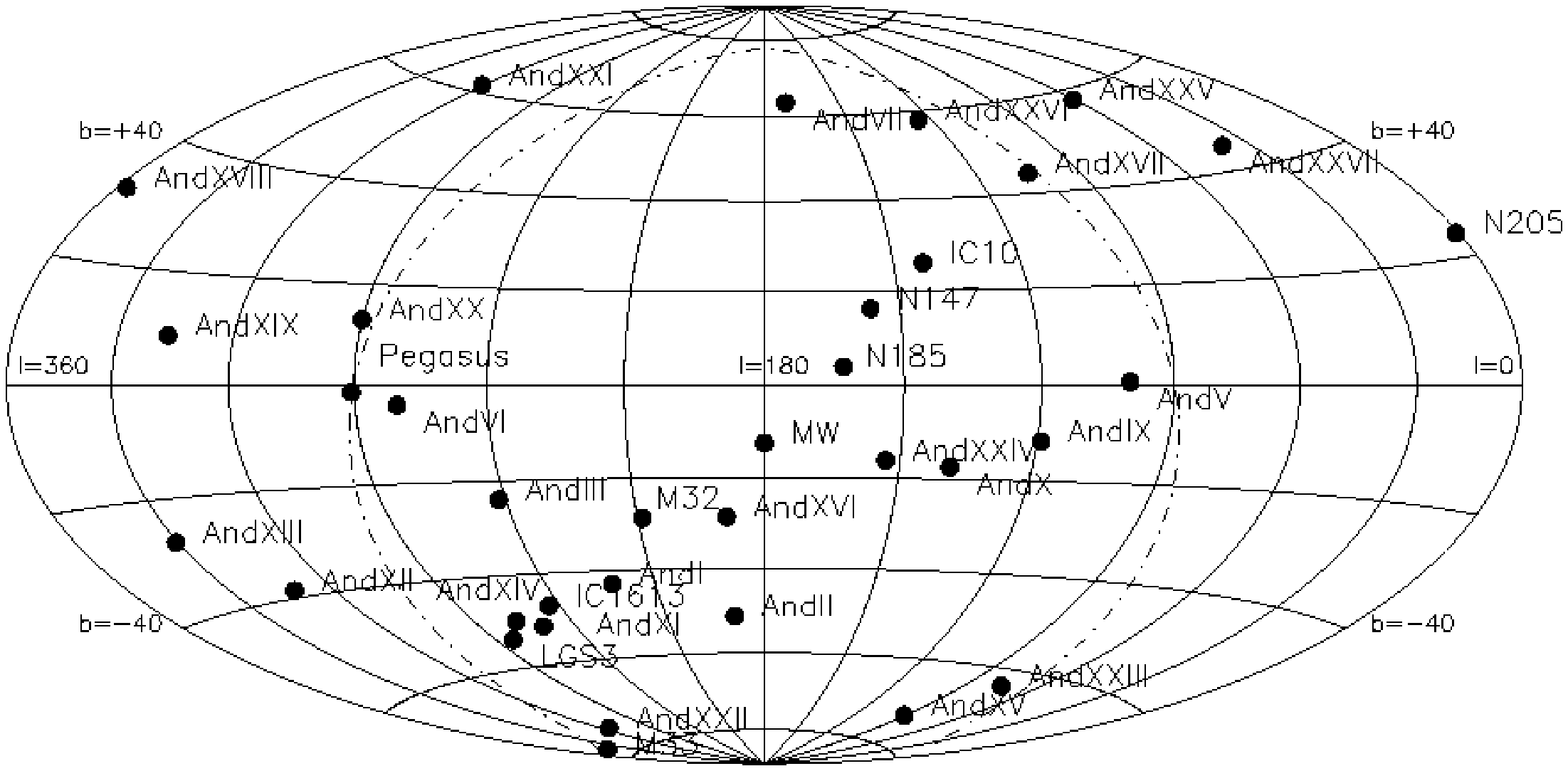}
\caption{\label{fig_dist} An Aitoff projection of the all-sky
distribution of the M31 satellite system as viewed from the center of
M31.  The co-ordinate system is defined using $l_{M31}$ measured
around the disk of M31 and locating the MW at $l_{M31} = 0^\circ$ and
$b_{m31} = -12.5^\circ$ to account for the inclination of the M31 disk
along the MW line-of-sight.  The dot-dash Great Circle centered on the
MW splits the sky into MW near-side and far-side.  With the extra M31
satellites now known the earlier apparently significant asymmetry in
the satellite distribution noted by \cite{mcconnachie06b} has now
largely disappeared.}
\end{center}
\end{figure*}

An interesting peculiarity in the M31 satellite distribution noted by
\cite{mcconnachie06b} was the tendency for the then known satellites
of M31 to be preferentially located on the MW side of M31 (14 out of
16).  With the discovery of 18 more M31 satellites since then it is
worth revisiting the satellite distribution again.  As before, we
define an M31-centric coordinate system such that the disk of M31
defines the galactic plane \ie latitude $b_{M31} = 0^\circ$, and such
that the MW lies at longitude $l_{M31} = 180^\circ$ and latitude
$b_{M31} = -12.5^\circ$.  Within this coordinate system the
distribution of M31 satellites, as would be seen from the center of
M31, is shown in the Aitoff projection of Fig.~\ref{fig_dist}.  The
outlined Great Circle has a pole centered on the MW position and
splits the M31 sky into those objects lying on the MW side within, and
on the opposite side to the MW, without.  Although there is still an
apparent bias, with 24 out of 34 satellites lying on the MW side of
M31, the offset in the line-of-sight distance distribution is much
less significant than that shown in figure 8. of
\cite{mcconnachie06b}.  With the additional satellites, and updated
information for the rest, the median offset toward the MW changes from
40~kpc to 21~kpc, and changes from a $3\sigma$ bias to less than a
$2\sigma$ event.


\begin{figure}
\begin{center}
\includegraphics[width=0.75\hsize,angle=270]{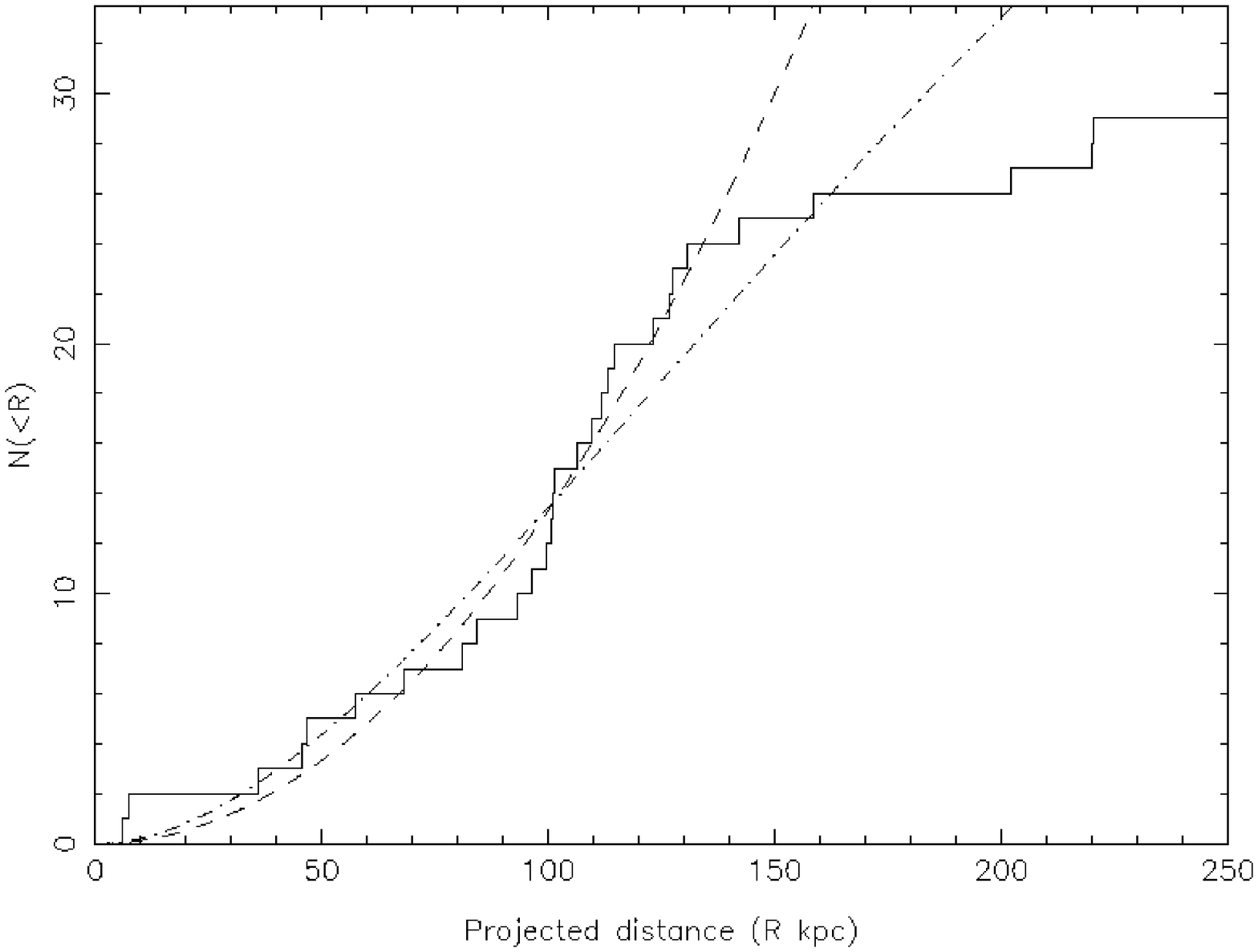}
\caption{\label{fig_cum} The cumulative distribution of the number of
satellite galaxies as a function of projected distance from M31. The
overlaid models are: a power law distribution of the form $\rho(r) \
\propto \ r^{-\alpha}$ with $\alpha = 1$ (dashed line); and an Einasto
density distribution with $\alpha = 0.678$ and $r_{-2} = 200\kpc$ (\eg
\citealt{springel08}; dot-dash line). Note particularly that the power
law model corresponds to a constant surface density of satellites and
that the clear break in the projected cumulative profile at the main
PAndAS survey limit of $\approx$150 kpc suggests a large number of
satellites remain to be discovered further out.  The cosmologically
motivated model does not provide a good fit to the observed surface
density distribution.}
\end{center}
\end{figure}

On the other hand, the overall spatial distribution of the satellites
of M31 reveals an unexpected result.  The projected (surface) spatial
distribution of satellites depicted in Fig.~\ref{fig_cum} highlights
the visual impression commented on previously from the two-dimensional
overview of satellite locations seen in Fig.~\ref{fig_map}.  Until the
main limit of the current PAndAS survey is reached at a projected
distance of $\approx$150 kpc from M31, there is an essentially uniform
surface density of satellite galaxies as first pointed out by
\cite{mcconnachie09}.  The turnover near 150 kpc in the projected
cumulative distribution corresponds precisely to the limit of the main
PAndAS survey.  Since we know from their three-dimensional
distribution that M31 satellites exist at distances up to at least
300--400 kpc from M31, it is quite likely that even with the current
PAndAS survey we are still seriously incomplete in our census of even
the relatively bright ($\rm{M}_V \simlt -7$) satellites.

In compiling this figure we have included all suspected Andromedan
satellites irrespective of their location relative to our currently
surveyed region.  The inclusion, or otherwise, of marginal Andromedan
members such as And~XVIII, makes little difference to the conclusion.
The surprising result is that out to our approximate survey limit at
$\approx$150 kpc in projection, the surface density of satellites is
essentially constant.  This corresponds to a three-dimensional radial
density distribution, $\rho(r) \propto r^{-1}$, a result seemingly in
conflict with cosmological simulations.  As a particular example of
one of these, we show in Fig.~\ref{fig_cum} a standard cosmological
model prediction of sub-halo density profiles based on an Einasto
radial density distribution with parameters $\alpha = 0.678$ and
$r_{-2} = 200\kpc$, taken from Figure~11 of \cite{springel08}.  Since
the normalization here is arbitrary, both overlaid models were
normalized to have the same mean projected surface density as observed
out to 100 kpc.  A constant surface density to $\approx$150 kpc in
projection provides a reasonable description of the observed satellite
population whereas the cosmologically motivated model has a much
steeper radial density fall-off and a correspondingly much flatter
predicted cumulative surface density distribution.

The present observed distribution of dwarf satellite galaxies will
undoubtedly have been influenced by their range of orbital properties
and evolutionary histories, with tidal effects preferentially
destroying systems closer to the nucleus of M31.  While we might
expect that the original distribution of recognizable satellites could
have been more centrally concentrated, the evolution over time of the
distribution, and properties, of the surviving systems is less
obvious.  Although detailed simulations of the distribution of dark
matter sub-halos have been made, the problem of accurately linking the
surviving sub-halos with the observed satellite systems still remains
(\eg \citealt{ludlow10}).

\section{Summary}

We have presented for the first time a panoramic view of metal-poor
stars in the M31 halo out to an average projected radius of
$\sim150\kpc$ using data obtained with the MegaPrime/MegaCam
wide-field camera on the CFHT during the PAndAS survey. Dwarf
satellite galaxies and tidal debris streams have been discovered out
to the edge of the survey suggesting that much more is to be found at
still larger radii.  The degree of sub-structure visible in the
overall map highlights the continuing growth and evolution of the
outer halo and disk of M31 and yields an insight into the accretion
process of an L$_*$ disk galaxy.

In the main part of the paper we have characterized the five most
recently discovered dwarf spheroidal galaxies uncovered by PAndAS,
Andromeda~XXIII - XXVII. They all contain stellar populations typical
of dSph galaxies, with no obvious sign of substructure, a relatively
narrow RGB and have mean metallicities ranging from [Fe/H] = $-1.7 \pm
0.2$~dex to [Fe/H] = $-1.9 \pm 0.2$~dex. Although the central surface
brightnesses of the galaxies are broadly similar ($\Sigma_{v,0} = 27.5
\pm 0.5 ~\rm{mag/arcsec}^2$, they vary in absolute magnitude
($\rm{M}_V = -7.1 \pm 0.5$ to -10.2$ \pm$ 0.5) and have very different
scale lengths ($\rm{R_h}$ = 230~pc to 1035~pc).  These additional
discoveries continue the trend whereby the brighter M31 dSphs have
significantly larger half-light radii than their MW counterparts.

\begin{itemize}
\item
\textbf{And~XXIII} is a bright ($\rm{M}_V = -10.2$) satellite galaxy
lying at the same line-of-sight distance as its host and $\sim
126$~kpc away from it. With a half-light radius of $\rm{r}_h =
1035$~pc it is one of the largest of M31's presently known satellites
and almost twice as extended as the largest known MW dSph satellite
(Fornax: $\rm{r}_h = 636$~pc, \citealt{irwin95}).

\item
\textbf{And~XXIV} and \textbf{And~XXVI}. These two galaxies resemble
the traditional dSphs belonging to the MW group in many ways. They
have small half-light-radii ($\rm{r}_h = 357$~pc and 232~pc) and are
relatively faint ($\rm{M}_V = -7.6 \pm 0.5$ and $-7.1 \pm 0.5$). Both
galaxies are very sparsely populated (see Figs~\ref{fig_spatial} and
\ref{fig_cmd}). Our line-of-sight distance measurements show that both
galaxies lie on the near side of M31 relative to the MW and, at a
distance of $600 \pm 33$~kpc, And~XXIV is one of the closest M31 dSphs
to us.

\item
\textbf{And~XXV} is characteristic of M31's dSph satellite system in
that it is large ($\rm{r}_h = 693$~pc) and bright ($\rm{M}_V =
-9.7$). It has the highest central surface brightness of the newly
discovered galaxies ($\Sigma_{V,0} = 27.1~\rm{mag/arcsec}^2$), still
not bright enough to have been found by previous shallower surveys
such as the SDSS.

\item
\textbf{And~XXVII} appears to be in the process of being tidally
disrupted by M31. It is embedded in a great Northern arc of stars
which loops around M31. The stream of stars surrounding the galaxy has
made it very difficult to extract a clean reference field to
characterize the properties of the foreground and/or background.  Of
the five galaxies presented in this paper, And~XXVII is furthest from
the MW ($\sim$827 $\pm$ 47~kpc) and the nearest to M31 ($\rm{R}_{M31}
\sim$86~kpc).

\end{itemize}

Finally, we revisited the spatial distribution of the overall M31
satellite population and demonstrated that the apparent bias in the
satellite location with respect to the MW noted by
\cite{mcconnachie06b} has gradually disappeared with the discovery of
more satellites.  However, in recompense, the newly discovered
satellites now allow an almost complete census out to a projected
distance of $\approx$150 kpc and show a surprising uniformity of their
surface density out to this limit.  This not only suggests that a
large fraction of the M31 satellite population remains to be
discovered but also that the three-dimensional radial density
distribution of M31 satellites varies approximately as $\rho(r)
\propto r^{-1}$ out to at least 150 kpc.  It will be enlightening to
compare this result with more detailed cosmological simulations based
on the estimated numbers of surviving satellites of $L_*$ galaxies.
Surveying to comparable depths to beyond the virial radius of M31
($\approx$250 kpc) is probably impractical with MegaCam but well
within the capabilities of the upcoming Hyper-Suprime Cam on Subaru
(\citealt{miyazaki06}) and would correspondingly provide a more
exacting test of cosmological predictions.

\acknowledgements We thank Sidney Van Den Bergh for a careful reading
of the manuscript and for his helpful comments on the content.  Thanks
also to the entire staff at CFHT for their great efforts and
continuing support throughout the PAndAS project.


\begin{appendix}

\section{ Fitting surface profiles}

We want to estimate the likelihood of observing the data points at position 
vectors $\{{\bf r_i}\} \ , \ i = 1,2,.....m$, where $f({\bf r})$ is the surface 
density profile of a given dwarf galaxy superimposed on a uniform local 
background.  Following the Press-Schechter formalism 
(Schechter \& Press 1976), imagine partitioning the observed region into 
``cells'' of surface area $\delta S$. Let the expected number of observed 
data points in cell $i$ be $\phi_i$ where
\begin{equation}
\phi_i = f({\bf r_i}) \ \delta S 
\end{equation}
then the probability of observing $x_i$ points in cell $i$ is given by a 
Poisson distribution
\begin{equation}
P(x_i) = e^{-\phi_i} \ {\phi_i^{x_i} \over x_i \  ! } 
\end{equation}
and therefore the likelihood function for the ensemble of data points in the
region is
\begin{equation}
L \ = \ \prod_i P(x_i) = \ \prod_i e^{-\phi_i} \ {\phi_i^{x_i} \over x_i \ ! } 
\end{equation}
Let $\delta S \rightarrow 0$, then $x_i = 1$  if a point is observed at this
location, or $x_i = 0$ if none are detected.   The likelihood function now 
becomes
\begin{equation}
L \ = \ \prod_i e^{-\phi_i} \ . \ \prod_i \phi_i \ e^{-\phi_i}
\end{equation}
where the first product is over empty cells and the second is over occupied
cells.  Taking natural logarithms this further simplifies to
\begin{equation}
ln\ L \ = \ \sum_i - \phi_i \ \ + \ \ \sum_{i=1}^m \  ln \ \phi_i
\end{equation}
where the first summation is now over all cells and the second only occupied
cells.  Since $\delta S \rightarrow 0$ we can replace the
first summation with a surface integral over the region such that
\begin{equation}
ln\ \ L = \ - \int_{S} f({\bf r}) \ dS \ + \ \sum_{i=1}^m \ ln \ f({\bf r_i}) 
\end{equation}
Unobservable regions due to gaps etc., are trivially included in the normalizing
integral by use of a window function, which of course has no impact on
the discrete summation of observed data points.

To illustrate the use of this method, we model the dSph as an ellipse with 
ellipticity $\epsilon$, position angle $\theta$ in a standard coordinate 
grid, and use an exponential profile of scale length $a$ along the major axis. 
In this case
\begin{equation}
f({\bf r_i}) \ = \ f_o \  e^{-r_i/a} + b
\end{equation}
where using standard coordinates $\Delta \xi$ and $\Delta \eta$ to denote
offsets from the center of the dSph \ie $\xi - \xi_o$ and $\eta - \eta_o$,
we have 
\begin{equation}
r_i^2 \ = \ (\Delta \xi_i \ cos\theta + \Delta \eta_i \ sin\theta)^2 + 
(-\Delta \xi_i \ sin\theta + \Delta \eta_i \ cos\theta)^2/(1-\epsilon)^2 
\end{equation}
If the area for analysis, $A$, covers sufficient ellipse scale lengths, and 
there are no gaps in the coverage, then
\begin{equation}
ln \ L = \ -2 \pi f_o \ a^2 (1-\epsilon) -bA + 
   \sum_{i=1}^m \ ln \ (f_o \  e^{-r_i/a} + b) 
\end{equation}
and for a given trial grid solution for $a, \epsilon, \theta$, and for a 
particular background $b$, the central density $f_o$ is defined by
\begin{equation}
{\partial ln \ L \over \partial f_o} \ = \ 0 \ = \ - 2 \pi a^2 (1-\epsilon)
 + \sum_{i=1}^m \ {e^{-r_i/a} \over f_o \  e^{-r_i/a} + b} 
\end{equation}
which has a straightforward iterative solution for $f_o$.  Likewise
the central coordinates of the dSph, in a coordinate system rotated for 
convenience to lie along the major and minor axes, $\xi_o'$, $\eta_o'$, are
defined by
\begin{equation}
{\partial ln \ L \over \partial \xi_o'} \ = \ 0 \ = \ 
 \sum_{i=1}^m \ {\xi_i' - \xi_o' \over r_i \ a} \ 
 { f_o \  e^{-r_i/a} \over f_o \  e^{-r_i/a} + b  }  
\end{equation}
which implies that 
\begin{equation}
\xi_o' \ = \ \sum_{i=1}^m \ { f_o \  e^{-r_i/a} \over f_o \  e^{-r_i/a} + b }
  \ {\xi_i' \over r_i} \ \  / \ \  
  \sum_{i=1}^m \ { f_o \  e^{-r_i/a} \over f_o \  e^{-r_i/a} + b } \ {1 \over r_i} 
\end{equation}
\ie a simple weighted sum over the coordinates, 
with a similar solution for $\eta_o'$.  

For numerical stability in solving the latter equation, we introduce a 
softening parameter in the core of the dSphs $r_i^2 \rightarrow r_i^2 + c^2$
where $c$ is fixed to be equivalent to a 20pc scale length.  This also makes 
some allowance for the often observed flattening of the inner profile of 
dSphs with respect to a pure exponential.

\end{appendix}


\end{document}